\documentclass[12pt]{article}

\usepackage{scicite}
\usepackage{times}

\usepackage{graphicx}
\usepackage{amsmath}
\usepackage{amssymb}
\usepackage{comment}
\usepackage{hyperref}

\newenvironment{sciabstract}{%
\begin{quote} \bf}
{\end{quote}}

\title{Universality, criticality and complexity of information propagation in social media}

\author
{Daniele Notarmuzi,$^{1}$ Claudio Castellano,$^{2}$ Alessandro  Flammini,$^{1}$\\ Dario Mazzilli,$^{1,3}$ Filippo Radicchi$^{1\ast}$\\
\\
\normalsize{$^{1}$Center for Complex Networks and Systems Research,}\\
\normalsize{Luddy School of Informatics, Computing, and Engineering}\\
\normalsize{Indiana University, Bloomington, Indiana 47408, USA}\\
\normalsize{$^{2}$Istituto dei Sistemi Complessi (ISC-CNR),}\\
\normalsize{Via dei Taurini 19, I-00185 Roma, Italy}\\
\normalsize{$^{3}$Centro Fermi}\\
\normalsize{Via Panisperna 89 A, Roma, Italy.}
\\
\normalsize{$^\ast$To whom correspondence should be addressed; E-mail:  filiradi@indiana.edu}
}

\date{}

\begin{document} 

\baselineskip24pt

\maketitle

\begin{sciabstract}

  Information avalanches in social media are typically studied in a similar fashion as avalanches of neuronal activity in the brain. Whereas a large body of literature reveals substantial agreement about the existence of a unique process characterizing neuronal activity across organisms, the dynamics of information in online social media is far less understood. Statistical laws of information avalanches are found in previous studies to be not robust
  across systems, and radically different processes are used to represent plausible driving mechanisms for information propagation. Here, we analyze almost 1 billion time-stamped events collected from a multitude of online platforms -- including Telegram, Twitter and Weibo -- over observation windows longer than 10 years to show that the propagation of information in social media is
  a universal and critical process.
  Universality arises from the observation of identical macroscopic patterns across platforms, irrespective of the details of the specific system at hand. Critical behavior is deduced from the power-law distributions, and corresponding hyperscaling relations, characterizing size and duration of  avalanches of information. 
  Neuronal activity may be modeled as a simple contagion process, where only a single exposure to activity may be sufficient for its diffusion. On the contrary, statistical testing on our data 
  indicates that a mixture of simple and complex contagion, where involvement of an individual requires exposure from multiple acquaintances, characterizes the propagation of information in social media.
  We show that the complexity of the process is correlated with the semantic content of the information that is propagated.  Conversational topics about music, movies and TV shows tend to propagate as simple contagion processes, whereas controversial discussions on political/societal themes obey the rules of complex contagion. 
\end{sciabstract}

\newpage


Social media have dramatically changed the way people produce, access and consume information~\cite{ahmad2010twitter}, 
and there is increasing evidence that online discussions have the potential to impact society in unprecedented ways~\cite{kwak2010twitter}~\footnote{Only in the past year,
we witnessed two emblematic examples. 
The public debate around the COVID-19 pandemic has been accompanied by the so-called {\it Infodemic} that is 
affecting the outcome of the vaccination campaign by increasing 
hesitancy~\cite{pierri2021impact,yang2020prevalence,yang2021covid}.
Also, online discussions in
the Reddit channel {\it r/wallstreetbets}
induced many individuals to buy GameStop shares
in opposition to the shorting operation 
carried out by hedge funds and professional 
investors. As a result, 
the market capital of the company displayed an increase of
more than \$22 billion in just a few days~\cite{phillips2021dumb}.}. 
It is not surprising therefore the renewed scientific interest to comprehend the mechanisms that drive information propagation.

Analyses of the propagation of information in social media reveals, at least qualitatively, similarities with other natural phenomena such as  
the firing of 
neurons~\cite{dalla2019modeling,beggs2003neuronal} and earthquakes~\cite{bak2002unified}. These are processes characterized by bursty activity patterns. Activity consists of point-like events in time, and bursts (or avalanches) of activity are defined as sequences of close-by events. Bursts are separated by long periods of low activity. Activity is characterized at the macroscopic level by the distributions $P(S)$ and $P(T)$ of the size $S$ and the duration $T$ of avalanches. Information propagation  can be studied considering the same observables~\cite{gleeson2014competition, barabasi2005origin, karsai2012universal, nishi2016reply, wegrzycki2017cascade, lerman2010information}. 
In real-word systems $P(S)$ and $P(T)$ have a power-law decay for large value of their argument,  i.e., $P(S) \sim S^{-\tau}$ and $P(T) \sim T^{-\alpha}$~\cite{munoz1999avalanche, dalla2019modeling, beggs2003neuronal, bak2002unified, karsai2012universal, onnela2010spontaneous, munoz2018colloquium}.
This property is interpreted as evidence of the system operating at, or in the vicinity of, a critical point. This statement is supported by the theory of absorbing phase transitions according to which, if the avalanche dynamics is at a critical point, then $P(S)$ and $P(T)$ must decay as power laws.  Further, in a process operating at criticality, the average size of avalanches with given duration must obey the hyperscaling relation $\langle S \rangle \sim T^{\gamma}$, with $\gamma = (\alpha-1)/(\tau -1)$~\cite{munoz1999avalanche, sethna2001crackling, colaiori2008exactly}. The specific value of the exponents $\tau$ and $\alpha$ typically differ for classes of systems. Their actual values are fundamental for the characterization of systems into universality classes, i.e., an ontology of processes with conceptual and practical relevance~\cite{odor2004universality}. 

There are systems for which strong evidence supporting the existence of universality classes suggests appropriate theoretical models and the microscopic mechanisms driving the dynamics.
For example, there is large agreement on the fact that neuronal activity in the brain is universal and critical~\cite{dalla2019modeling, beggs2003neuronal,fontenele2019criticality,beggs2012being,haldeman2005critical,friedman2012universal}. Universality is the notion  that nearly identical avalanche statistics are observed for a multitude of organisms. Criticality instead refers to the fact that avalanche statistics are characterized by algebraic distributions. In particular, the critical exponent values  are those of the universality class of the mean-field branching process (BP), i.e., $\tau = 3/2$ and $\alpha = 2$~\cite{watson1875probability, harris1963theory, liggett2012interacting}. The finding informs us about the mechanism that drives the unfolding of an avalanche of neuronal activity in the brain. Neurons influence each other according to a simple ``contagion process,'' where only a single exposure to an active neuron may be sufficient to trigger the activity of another. As a result, activity propagates from neuron to neuron as the avalanche unfolds.

Where information propagation (in general, and in online social media) is concerned, the issue of the existence of well-defined universality classes is far from settled. 
Existing analyses typically study data collected from a single source and over short observation windows.  It is often found that distributions of avalanche size and duration obey power laws, but the estimated values of the exponents are not the same across studies: $\tau$ values range between $\tau \simeq 2$ and $\tau \simeq 4$~\cite{sreenivasan2016information,zhou2021survey,wegrzycki2017cascade,nishi2016reply,cao2017deephawkes}, whereas $\alpha \simeq 3.6$~\cite{oliveiraa2019diffusion}
or $\alpha \simeq 2.5$~\cite{bild2015aggregate,weng2012competition}.
Also, empirical studies reporting on correlations between size and duration of avalanches do not observe a power law at all~\cite{gleeson2016effects,szabo2010predicting}. 
These different results might be ascribed to multiple operative
definitions of avalanches, which can be given in terms of hashtags time series 
(TS)~\cite{sreenivasan2016information,gleeson2016effects}
as well as reply trees or retweet chains~\cite{nishi2016reply,li2019infectivity,cao2017deephawkes}. Further, regardless of the
definition, the temporal resolution can affect the avalanche 
distribution~\cite{karsai2012universal,notarmuzi2021percolation}.
As a consequence of the variability in the distributions inferred, uncertainty about representative theoretical models remains.
Finally, empirical evidence and theoretical support for microscopic mechanisms that may drive the propagation of information in social media are inconclusive. Stemming from the apparent similarity between the spreading of disease and information, a widely accepted paradigm is that information diffuses according to a simple contagion process~\cite{gleeson2014competition,o2020quantifying,sreenivasan2016information,nishi2016reply,crane2008robust,gleeson2016effects}. Simple contagion is at the core of many theoretical models of information propagation used in the literature, all displaying critical properties of the BP universality class~\cite{radicchi2020classes}. However, there are quite a few studies in favor of the complex contagion paradigm~\cite{weng2014predicting,vasconcelos2019consensus,state2015diffusion,hodas2014simple}. As originally introduced by Centola and Macy, in a complex contagion process the involvement of an individual in the propagation of information requires exposure from multiple acquaintances~\cite{centola2007complex}. 
Distinguishing between simple and complex contagion and, possibly, how they
coexist within the same population~\cite{guilbeault2018complex}, is
fundamental to 
understand the spreading of (mis)information in online social media~\cite{weng2014predicting,romero2011differences}. 
Complex contagion is exemplified by some models, such as
the Linear Threshold Model and the Random Field Ising Model (RFIM)~\cite{dodds2005generalized, sethna2001crackling}.


In this work,
we perform a large-scale study of (hash)tags TS from Twitter, Telegram, Weibo, Parler, StackOverflow and Delicious  
(see Supplemental Material (SM) A, B for details about the data sets). We consider a total of $206,972,692$ TS, cumulatively consisting of $905,377,009$ events, collected over periods ranging up to $10$ years.
The Twitter data, collected specifically for this work, 
are fully available together with codes to reproduce the results of
this paper~\cite{TWTdata}.
To define avalanches in a principled fashion we adopt the approach inspired by percolation theory proposed in 
Ref.~\cite{notarmuzi2021percolation}. We provide evidence that social media share universal statistics of avalanches that are well described by power-law distributions. At the aggregate level, 
each social media displays a critical behavior that is compatible with the RFIM,
indicating that, plausibly,
information propagates in social media according to a complex contagion process.
Second, we develop a novel statistical technique able to determine 
the level of criticality and complexity of individual TS.
We find that nearly 20\% of the TS are less than $5\%$ away from 
criticality. These account for 53\% of all events in our data sets. Also, 
we find that about $50\%$ of the individual TS are better explained in 
terms of a complex rather than a simple contagion process. A qualitative 
analysis of the most popular hashtags suggests that information concerning 
conversational topics, e.g., music or TV shows, spreads according to the 
rules of simple contagion, whereas information concerning
political/societal controversies
shows signatures of an underlying complex contagion process.


The operational definition of an avalanche depends on the value of the parameter $\Delta$, the minimal time separation between two consecutive events belonging to distinct avalanches.  
A proper choice $\Delta^*$ of the time resolution $\Delta$ for the specific data set at hand is necessary to avoid significant distortion in the resulting avalanche statistics. This statement is true for synthetic TS generated by temporal point processes~\cite{notarmuzi2021percolation}, but also
for the empirical TS analyzed in this paper (see SM D, I for details).
To determine the value of  $\Delta^*$ we take advantage of the principled method developed in Ref.~\cite{notarmuzi2021percolation} which identifies $\Delta^*$ as the critical point of a one-dimensional percolation model. Results are presented in Fig.~1. Values of  $\Delta^*$ for each data set are reported in the SM D; 
they vary substantially across data sets, 
from $\Delta^* \simeq 1,500$ s for Twitter to $\Delta^* \simeq 30,000$ s for Telegram (Fig.~1B). 

Once the time resolution is rescaled according to $\Delta \to \Delta/\Delta^*$, the curves of percolation strength relative to different data sets exhibit a nearly identical quantitative behavior. This fact
suggests the possibility of seeing the propagation of information in social media as a universal process, with $\Delta^*$ representing the natural resolution for observing information avalanches. 
Fig.~2A and ~2B show the distributions of avalanche size and duration obtained by setting $\Delta = \Delta^*$. Fig.~2C shows the relation between size and duration.  The collapse of curves relative to different data sets on a single curve hints once more to processes belonging to the same universality class.


The avalanche statistics of Figs.~2A-C seems well described by power laws, indicating that the underlying process is (nearly) critical, and that its universality class can be identified by estimating the value of the critical exponents $\tau$, $\alpha$, and $\gamma$~\cite{odor2004universality}. We rely on maximum likelihood estimation for $\tau$ and $\alpha$~\cite{clauset2009power}; linear regression on the logarithm of the relation $\langle S \rangle \sim T^{\gamma}$ is used to estimate $\gamma$. Results are reported in Fig.~2D, see SM G for details. The estimated exponent $\hat{\tau}$ 
is compatible with the one of the mean-field RFIM universality class, i.e., $\tau=9/4$~\cite{sethna2001crackling}. The compatibility of avalanche statistics with those of a homogeneous mean-field model is not surprising given that in some social media there is no underlying network among users and in the others there are mechanisms for the propagation of information that bypass it.
There is an apparent mismatch between our estimates
$\hat{\alpha}$ and $\hat{\gamma}$ and the RFIM predictions $\alpha=7/2$ and $\gamma=2$. The mismatch can be theoretically explained by the peculiar shape of the scaling function characterizing the distribution of avalanche duration, which affects also the estimate of
$\hat{\gamma}$~\cite{di2017simple}.
Difficulties in observing the asymptotic exponents of the RFIM  due to the effect of the scaling functions emerge also in numerical simulations of the RFIM and are well known~\cite{sethna2001crackling}. 

The proximity of exponents estimated across data sets points to the existence of a genuine and distinctive universality class for information propagation in social media. In particular, this class seems to be different from that of BP often invoked as representative in phenomena related to information diffusion.
If we repeat, for example, the same analysis on TS describing activity in very different types of systems, e.g., brain networks and earthquakes, avalanche duration and size still decay in a power-law fashion, but with radically different exponent values, see SM H for details. In particular, for neuronal avalanches in the brain we recover exponents compatible with the BP universality class.

To assess if the statistical properties obtained on aggregate data are representative of individual TS, we develop a maximum likelihood method to fit TS against BP and RFIM models. The technique is inspired by the work of Ref.~\cite{clauset2009power}, see SM K for details.  The method allows us to perform three different tests. First, it establishes the regime of a TS, depending on how the best estimate of the branching ratio parameter $n$ compares to the critical value $n_c=1$ for BP, or how the best estimate of the disorder parameter $R$ compares to the critical value $R_c= \sqrt{2/\pi} \simeq 0.8$ for RFIM. Second, it evaluates the goodness of the individual fits via their $p$-values. Similarly to the prescription of Ref.~\cite{clauset2009power}, we set the threshold for statistical significance equal to $p =0.1$. We verified, however, that the outcome of the analysis is not greatly affected by the choice of the threshold value, see SM O. Third, it establishes whether a TS is better modeled by BP or RFIM  by comparing their likelihood. 

Results of our analysis are reported in Figs.~3 and ~4. Our method is applied only to TS including avalanches that contain 
at least two avalanches larger than  $S_{min} = 10$.
Tests of robustness for different $S_{min}$ values are reported in the SM O. In all systems under analysis, we find that the best fitting parameter assumes values over a broad range, encompassing a large portion  of the subcritical phase, as well as the critical point of the models (Figs.~3A and 3B). 
The individual-level analysis 
confirms the results obtained for the aggregate data. The majority of events belongs to a minority of TS giving rise to 
 the largest avalanches. As a consequence, the 
large-scale behaviour of each system is mainly determined by those few TS that are  fitted in a 
 narrow region of the parameter space  close to the critical point for both BP and RFIM  (insets of Figs.~3A and~3B). Also, our tests indicate that the vast majority of TS are  well described by at least one of the two models (Fig.~4A). Model selection indicates that 
individual TS are divided in two nearly equally populated classes, one better described 
by  BP  and the other by  RFIM  (Fig.~4A). Simple and complex
contagion thus coexist in social media,
with only a mild dominance of complex over simple contagion (Fig.~3C).
These results are not incompatible with the aggregate avalanche statistics (Fig.~2). 
Fig.~3D shows that critical TS that belong to the class of complex contagion
display power-law scaling compatible with the RFIM $\tau$ exponent. 
Also, the critical TS that the fitting procedure attributes to the BP class show a neat crossover to RFIM scaling for large avalanches.
The mixture  
produces a universal distribution that is overall more compatible with the RFIM universality class rather than the BP class (Fig.~2C).

In summary, we revealed that temporal patterns characterizing 
bursts of activity in online social media are universal,
thus they should be
ascribed to 
mechanisms that are so basic that underlie information diffusion in all social
media platforms.
Also, in contrast with the vast majority of previous studies 
where purely diffusive models have been considered~\cite{radicchi2020classes}, we showed 
that information propagation in social media is often better described by a complex contagion dynamics.
Complex contagion is here exemplified by the RFIM, an agent-based model of 
activation originally formulated to describe the para-to-ferromagnetic phase transition in metals~\cite{sethna2001crackling}. Recast in language proper to the description of 
information propagation~\cite{michard2005theory}, RFIM prescribes that  
each agent (i) has a personal opinion, (ii) is subject to the social 
influence exerted by the agents she interacts with, and (iii) 
is also driven by an external force representing the public 
information about exogenous events. These appear reasonable
assumptions for  modeling  many realistic discussions  happening in social media. Fig.~4 shows the 30 most popular Twitter hashtags identified by
our method either in the simple or in the complex contagion classes.
In the category of simple contagion,
we find 
conversational topics, mostly related to
music or cinema/TV shows. 
Hashtags belonging to the class of  complex contagion  
display either periodic patterns or are related to political/controversial
themes. 
This qualitative picture
fits with previous studies that have explicitly focused
on the semantic meaning of different hashtags in 
Twitter~\cite{romero2011differences}. 
For both classes of information avalanches, we inferred the dynamics 
underlying their generation as critical, a fact that provides
theoretical ground for the surprising but remarkable robustness of our findings. Our results pave the way for future research about both descriptive
theories and data-driven predictive models. The presence of a large portion of social media content that acquires popularity via complex contagion
dynamics calls for a reconsideration of predictive algorithms relying on
the temporal characteristics of the signal only, 
because these algorithms
often neglect the semantics of hashtags and, even more frequently, topological features of their
propagation~\cite{kobayashi2016tideh,zhao2015seismic,matsubara2012rise,rizoiu2017expecting,haimovich2020scalable}. 
Both aspects are important for a successful discrimination 
between information propagating as a simple or complex contagion
process~\cite{romero2011differences,weng2014predicting}. 
We argue that the distinction between these truly
different mechanisms is 
fundamental for the development of
novel theoretical and data-driven 
approaches. We speculate that our results extend beyond the six platforms considered here. If so, there must be a mechanism that explains the 
universality  shown by the data, involving a critical dynamics that 
is independent of the peculiarities implemented in the individual platforms.
Understanding where this mechanism is rooted in and how to exploit this mechanism
for the prediction of the propagation of information in online social media remain open challenges for future research.~\nocite{osome,twitterdeca,baumgartner2020pushshift,aliapoulios2021early,fu2013assessing,basile_topical_2015,TLGdata,PARLdata,WEIdata,SOdata,DELdata,timme2016criticality,ito2014large,litke2004does,lawlor2018linear,JAPdata,CALdata,EURdata,grunthal2013share,RHDCdata,MSOSdata,MPCdata,stauffer2018introduction}

\newpage

\bibliography{biblio}

\begin{thebibliography}{10}

\bibitem{ahmad2010twitter}
A.~N. Ahmad, {\it Journal of media practice\/} {\bf 11}, 145 (2010).

\bibitem{kwak2010twitter}
H.~Kwak, C.~Lee, H.~Park, S.~Moon, {\it Proceedings of the 19th international
  conference on World wide web\/} (2010), pp. 591--600.

\bibitem{pierri2021impact}
F.~Pierri, {\it et~al.\/}, {\it arXiv preprint arXiv:2104.10635\/}  (2021).

\bibitem{yang2020prevalence}
K.-C. Yang, C.~Torres-Lugo, F.~Menczer, {\it arXiv preprint arXiv:2004.14484\/}
   (2020).

\bibitem{yang2021covid}
K.-C. Yang, {\it et~al.\/}, {\it Big Data \& Society\/} {\bf 8},
  20539517211013861 (2021).

\bibitem{phillips2021dumb}
M.~Phillips, T.~Lorenz, {\it The New York Times. https://www. nytimes.
  com/2021/01/27/business/gamestop-wall-street-bets. html\/}  (2021).

\bibitem{dalla2019modeling}
L.~Dalla~Porta, M.~Copelli, {\it PLoS computational biology\/} {\bf 15},
  e1006924 (2019).

\bibitem{beggs2003neuronal}
J.~M. Beggs, D.~Plenz, {\it Journal of neuroscience\/} {\bf 23}, 11167 (2003).

\bibitem{bak2002unified}
P.~Bak, K.~Christensen, L.~Danon, T.~Scanlon, {\it Physical Review Letters\/}
  {\bf 88}, 178501 (2002).

\bibitem{gleeson2014competition}
J.~P. Gleeson, J.~A. Ward, K.~P. O'sullivan, W.~T. Lee, {\it Physical review
  letters\/} {\bf 112}, 048701 (2014).

\bibitem{barabasi2005origin}
A.-L. Barabasi, {\it Nature\/} {\bf 435}, 207 (2005).

\bibitem{karsai2012universal}
M.~Karsai, K.~Kaski, A.-L. Barab{\'a}si, J.~Kert{\'e}sz, {\it Scientific
  reports\/} {\bf 2}, 1 (2012).

\bibitem{nishi2016reply}
R.~Nishi, {\it et~al.\/}, {\it Social Network Analysis and Mining\/} {\bf 6},
  26 (2016).

\bibitem{wegrzycki2017cascade}
K.~Wegrzycki, P.~Sankowski, A.~Pacuk, P.~Wygocki, {\it Proceedings of the 26th
  International Conference on World Wide Web\/} (2017), pp. 569--576.

\bibitem{lerman2010information}
K.~Lerman, R.~Ghosh, {\it Proceedings of the International AAAI Conference on
  Web and Social Media\/} (2010), vol.~4.

\bibitem{munoz1999avalanche}
M.~A. Munoz, R.~Dickman, A.~Vespignani, S.~Zapperi, {\it Physical Review E\/}
  {\bf 59}, 6175 (1999).

\bibitem{onnela2010spontaneous}
J.-P. Onnela, F.~Reed-Tsochas, {\it Proceedings of the National Academy of
  Sciences\/} {\bf 107}, 18375 (2010).

\bibitem{munoz2018colloquium}
M.~A. Munoz, {\it Reviews of Modern Physics\/} {\bf 90}, 031001 (2018).

\bibitem{sethna2001crackling}
J.~P. Sethna, K.~A. Dahmen, C.~R. Myers, {\it Nature\/} {\bf 410}, 242 (2001).

\bibitem{colaiori2008exactly}
F.~Colaiori, {\it Advances in Physics\/} {\bf 57}, 287 (2008).

\bibitem{odor2004universality}
G.~{\'O}dor, {\it Reviews of Modern Physics\/} {\bf 76}, 663 (2004).

\bibitem{fontenele2019criticality}
A.~J. Fontenele, {\it et~al.\/}, {\it Physical review letters\/} {\bf 122},
  208101 (2019).

\bibitem{beggs2012being}
J.~M. Beggs, N.~Timme, {\it Frontiers in physiology\/} {\bf 3}, 163 (2012).

\bibitem{haldeman2005critical}
C.~Haldeman, J.~M. Beggs, {\it Physical review letters\/} {\bf 94}, 058101
  (2005).

\bibitem{friedman2012universal}
N.~Friedman, {\it et~al.\/}, {\it Physical review letters\/} {\bf 108}, 208102
  (2012).

\bibitem{watson1875probability}
H.~W. Watson, F.~Galton, {\it The Journal of the Anthropological Institute of
  Great Britain and Ireland\/} {\bf 4}, 138 (1875).

\bibitem{harris1963theory}
T.~E. Harris, {\it et~al.\/}, {\it The theory of branching processes\/}, vol.~6
  (Springer Berlin, 1963).

\bibitem{liggett2012interacting}
T.~M. Liggett, {\it Interacting particle systems\/}, vol. 276 (Springer Science
  \& Business Media, 2012).

\bibitem{sreenivasan2016information}
S.~Sreenivasan, K.~S. Chan, A.~Swami, G.~Korniss, B.~K. Szymanski, {\it IEEE
  Transactions on Network Science and Engineering\/} {\bf 4}, 120 (2016).

\bibitem{zhou2021survey}
F.~Zhou, X.~Xu, G.~Trajcevski, K.~Zhang, {\it ACM Computing Surveys (CSUR)\/}
  {\bf 54}, 1 (2021).

\bibitem{cao2017deephawkes}
Q.~Cao, H.~Shen, K.~Cen, W.~Ouyang, X.~Cheng, {\it Proceedings of the 2017 ACM
  on Conference on Information and Knowledge Management\/} (2017), pp.
  1149--1158.

\bibitem{oliveiraa2019diffusion}
D.~F. Oliveiraa, K.~S. Chana, {\it Information \& Security\/} {\bf 43}, 362
  (2019).

\bibitem{bild2015aggregate}
D.~R. Bild, Y.~Liu, R.~P. Dick, Z.~M. Mao, D.~S. Wallach, {\it ACM Transactions
  on Internet Technology (TOIT)\/} {\bf 15}, 1 (2015).

\bibitem{weng2012competition}
L.~Weng, A.~Flammini, A.~Vespignani, F.~Menczer, {\it Scientific Reports\/}
  {\bf 2}, 335 (2012). Number: 1 Publisher: Nature Publishing Group.

\bibitem{gleeson2016effects}
J.~P. Gleeson, K.~P. O'Sullivan, R.~A. Ba{\~n}os, Y.~Moreno, {\it Physical
  Review X\/} {\bf 6}, 021019 (2016).

\bibitem{szabo2010predicting}
G.~Szabo, B.~A. Huberman, {\it Communications of the ACM\/} {\bf 53}, 80
  (2010).

\bibitem{li2019infectivity}
W.~Li, S.~J. Cranmer, Z.~Zheng, P.~J. Mucha, {\it PloS one\/} {\bf 14},
  e0214453 (2019).

\bibitem{notarmuzi2021percolation}
D.~Notarmuzi, C.~Castellano, A.~Flammini, D.~Mazzilli, F.~Radicchi, {\it
  Physical Review E\/} {\bf 103}, L020302 (2021).

\bibitem{o2020quantifying}
J.~D. O'Brien, A.~Aleta, Y.~Moreno, J.~P. Gleeson, {\it arXiv preprint
  arXiv:2001.09490\/}  (2020).

\bibitem{crane2008robust}
R.~Crane, D.~Sornette, {\it Proceedings of the National Academy of Sciences\/}
  {\bf 105}, 15649 (2008).

\bibitem{radicchi2020classes}
F.~Radicchi, C.~Castellano, A.~Flammini, M.~A. Mu{\~n}oz, D.~Notarmuzi, {\it
  Physical Review Research\/} {\bf 2}, 033171 (2020).

\bibitem{weng2014predicting}
L.~Weng, F.~Menczer, Y.-Y. Ahn, {\it Proceedings of the International AAAI
  Conference on Web and Social Media\/} (2014), vol.~8.

\bibitem{vasconcelos2019consensus}
V.~V. Vasconcelos, S.~A. Levin, F.~L. Pinheiro, {\it Journal of the Royal
  Society Interface\/} {\bf 16}, 20190196 (2019).

\bibitem{state2015diffusion}
B.~State, L.~Adamic, {\it Proceedings of the 18th ACM Conference on Computer
  Supported Cooperative Work \& Social Computing\/} (2015), pp. 1741--1750.

\bibitem{hodas2014simple}
N.~O. Hodas, K.~Lerman, {\it Scientific reports\/} {\bf 4}, 1 (2014).

\bibitem{centola2007complex}
D.~Centola, M.~Macy, {\it American journal of Sociology\/} {\bf 113}, 702
  (2007).

\bibitem{guilbeault2018complex}
D.~Guilbeault, J.~Becker, D.~Centola, {\it Complex spreading phenomena in
  social systems\/} pp. 3--25 (2018).

\bibitem{romero2011differences}
D.~M. Romero, B.~Meeder, J.~Kleinberg, {\it Proceedings of the 20th
  international conference on World wide web\/} (2011), pp. 695--704.

\bibitem{dodds2005generalized}
P.~S. Dodds, D.~J. Watts, {\it Journal of theoretical biology\/} {\bf 232}, 587
  (2005).

\bibitem{TWTdata}
D.~Notarmuzi, C.~Castellano, A.~Flammini, D.~Mazzilli, F.~Radicchi,
  \url{https://github.com/DaniMuzi/SocialMedia} (2021).

\bibitem{clauset2009power}
A.~Clauset, C.~R. Shalizi, M.~E. Newman, {\it SIAM review\/} {\bf 51}, 661
  (2009).

\bibitem{di2017simple}
S.~di~Santo, P.~Villegas, R.~Burioni, M.~A. Mu{\~n}oz, {\it Physical Review
  E\/} {\bf 95}, 032115 (2017).

\bibitem{michard2005theory}
Q.~Michard, J.-P. Bouchaud, {\it The European Physical Journal B-Condensed
  Matter and Complex Systems\/} {\bf 47}, 151 (2005).

\bibitem{kobayashi2016tideh}
R.~Kobayashi, R.~Lambiotte, {\it Proceedings of the International AAAI
  Conference on Web and Social Media\/} (2016), vol.~10.

\bibitem{zhao2015seismic}
Q.~Zhao, M.~A. Erdogdu, H.~Y. He, A.~Rajaraman, J.~Leskovec, {\it Proceedings
  of the 21th ACM SIGKDD international conference on knowledge discovery and
  data mining\/} (2015), pp. 1513--1522.

\bibitem{matsubara2012rise}
Y.~Matsubara, Y.~Sakurai, B.~A. Prakash, L.~Li, C.~Faloutsos, {\it Proceedings
  of the 18th ACM SIGKDD international conference on Knowledge discovery and
  data mining\/} (2012), pp. 6--14.

\bibitem{rizoiu2017expecting}
M.-A. Rizoiu, {\it et~al.\/}, {\it Proceedings of the 26th International
  Conference on World Wide Web\/} (2017), pp. 735--744.

\bibitem{haimovich2020scalable}
D.~Haimovich, D.~Karamshuk, T.~J. Leeper, E.~Riabenko, M.~Vojnovic, {\it arXiv
  preprint arXiv:2009.02092\/}  (2020).

\bibitem{osome}
I.~University, {OS}o{M}e, {O}bservatory on {S}ocial {M}edia,
  \url{https://osome.iu.edu} (2020).

\bibitem{twitterdeca}
Twitter, {D}ecahose stream,
  \url{https://developer.twitter.com/en/docs/twitter-api/v1/tweets/sample-realtime/overview/decahose}.

\bibitem{baumgartner2020pushshift}
J.~Baumgartner, S.~Zannettou, M.~Squire, J.~Blackburn, {\it arXiv preprint
  arXiv:2001.08438\/}  (2020).

\bibitem{aliapoulios2021early}
M.~Aliapoulios, {\it et~al.\/}, {\it arXiv preprint arXiv:2101.03820\/}
  (2021).

\bibitem{fu2013assessing}
K.-w. Fu, C.-h. Chan, M.~Chau, {\it IEEE Internet Computing\/} {\bf 17}, 42
  (2013).

\bibitem{basile_topical_2015}
V.~Basile, S.~Peroni, F.~Tamburini, F.~Vitali, {\it J. Information Science\/}
  {\bf 41}, 486 (2015).

\bibitem{TLGdata}
J.~Baumgartner, S.~Zannettou, M.~Squire, J.~Blackburn,
  \url{https://zenodo.org/record/3607497#.YRu-4tMza-s} (2020).

\bibitem{PARLdata}
M.~Aliapoulios, {\it et~al.\/},
  \url{https://zenodo.org/record/4442460#.YRu_WtMza-s} (2021).

\bibitem{WEIdata}
K.-w. Fu, Weiboscope open data. (dataset),
  \url{https://hub.hku.hk/cris/dataset/dataset107483} (2017).

\bibitem{SOdata}
Link to stackoverflow data,
  \url{https://archive.org/download/stackexchange/stackoverflow.com-Posts.7z}.

\bibitem{DELdata}
V.~Basile, \url{http://valeriobasile.github.io/delicious/} (2015).

\bibitem{timme2016criticality}
N.~M. Timme, {\it et~al.\/}, {\it Frontiers in physiology\/} {\bf 7}, 425
  (2016).

\bibitem{ito2014large}
S.~Ito, {\it et~al.\/}, {\it PloS one\/} {\bf 9}, e105324 (2014).

\bibitem{litke2004does}
A.~Litke, {\it et~al.\/}, {\it IEEE Transactions on Nuclear Science\/} {\bf
  51}, 1434 (2004).

\bibitem{lawlor2018linear}
P.~N. Lawlor, M.~G. Perich, L.~E. Miller, K.~P. Kording, {\it Journal of
  computational neuroscience\/} {\bf 45}, 173 (2018).

\bibitem{JAPdata}
\url{http://wwweic.eri.u-tokyo.ac.jp/CATALOG/junec/monthly.html}.

\bibitem{CALdata}
California earthquakes (dataset),
  \url{https://www.conservation.ca.gov/cgs/Documents/Melange/cgs2000_fnl.txt}.

\bibitem{EURdata}
G.~Gr{\"u}nthal, R.~Wahlstr{\"o}m, D.~Stromeyer, Data taken from sheec
  1900-2006 (gr{\"u}nthal et al., 2013) .,
  \url{https://www.gfz-potsdam.de/sheec/} (2013).

\bibitem{grunthal2013share}
G.~Gr{\"u}nthal, R.~Wahlstr{\"o}m, D.~Stromeyer, {\it Journal of seismology\/}
  {\bf 17}, 1339 (2013).

\bibitem{RHDCdata}
N.~M. Timme, {\it et~al.\/}, Dissociated cutlures of rat's hippocampal cells
  (dataset), \url{https://crcns.org/data-sets/hc/hc-8} (2016).

\bibitem{MSOSdata}
S.~Ito, {\it et~al.\/}, Spontaneous spiking activity of hundreds of neurons in
  mouse somatosensory cortex slice cultures recorded using a dense 512
  electrode array, \url{http://dx.doi.org/10.6080/K07D2S2F} (2016).

\bibitem{MPCdata}
M.~G. Perich, P.~N. Lawlor, K.~P. Kording, L.~E. Miller, Extracellular neural
  recordings from macaque primary and dorsal premotor motor cortex during a
  sequential reaching task., \url{http://dx.doi.org/10.6080/K0FT8J72} (2018).

\bibitem{stauffer2018introduction}
D.~Stauffer, A.~Aharony, {\it Introduction to percolation theory\/} (CRC press,
  2018).

\end{thebibliography}

\bibliographystyle{Science}

\section*{Acknowledgments}

F.R. acknowledges support from the National Science Foundation (CMMI-1552487). D.N. was partially funded by the National Science Foundation NRT grant 1735095.
Any opinions, findings, and conclusions or recommendations expressed in this work are those of the author(s) and do not necessarily reflect the views of the National Science Foundation.



\clearpage

\begin{figure}[htp]
    \centering
    \includegraphics[width=0.8\textwidth]{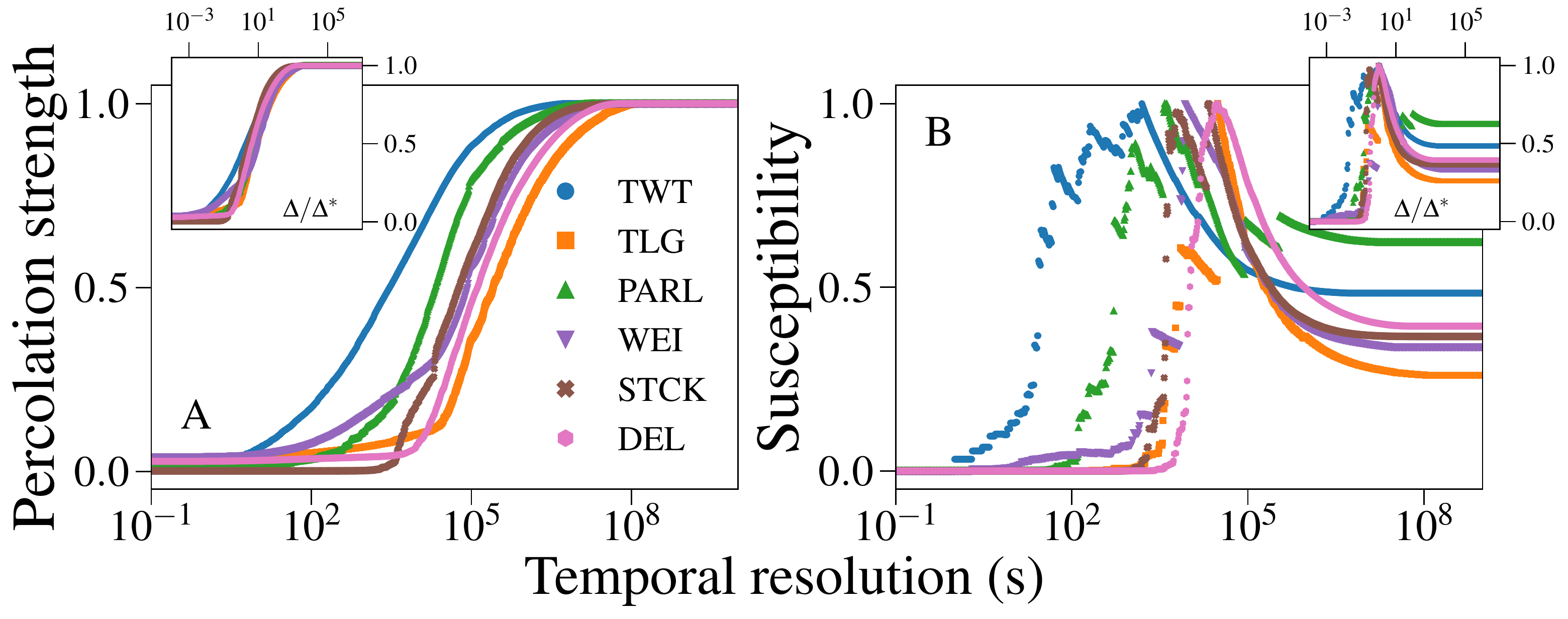}
\end{figure}

\noindent {\bf Fig. 1.} 
{\bf Universality of information propagation in online social media.}
A) In the main panel, we show the percolation strength as a function of the temporal resolution $\Delta$. Different colors/symbols refer to different social media: Twitter (TWT), Telegram (TLG), Parler (PARL), Weibo (WEI), Stack Overflow (STCK), and  Delicious (DEL). In the inset, we plot the same data as in the main panel, but with the horizontal axis rescaled as $\Delta \to \Delta/\Delta^*$.  
B) In the main panel, we plot the susceptibility as a function of the time resolution for the same data as in A. 
The optimal  resolution $\Delta^*$ is identified as the location of the peak  of the susceptibility. In the inset, we plot the same data as in the main panel, but with the rescaling  $\Delta \to \Delta/\Delta^*$. For the sake of
comparison, each curve has been normalized to its maximum.

\newpage
\begin{figure}[htp]
    \centering
    \includegraphics[width=0.9\textwidth]{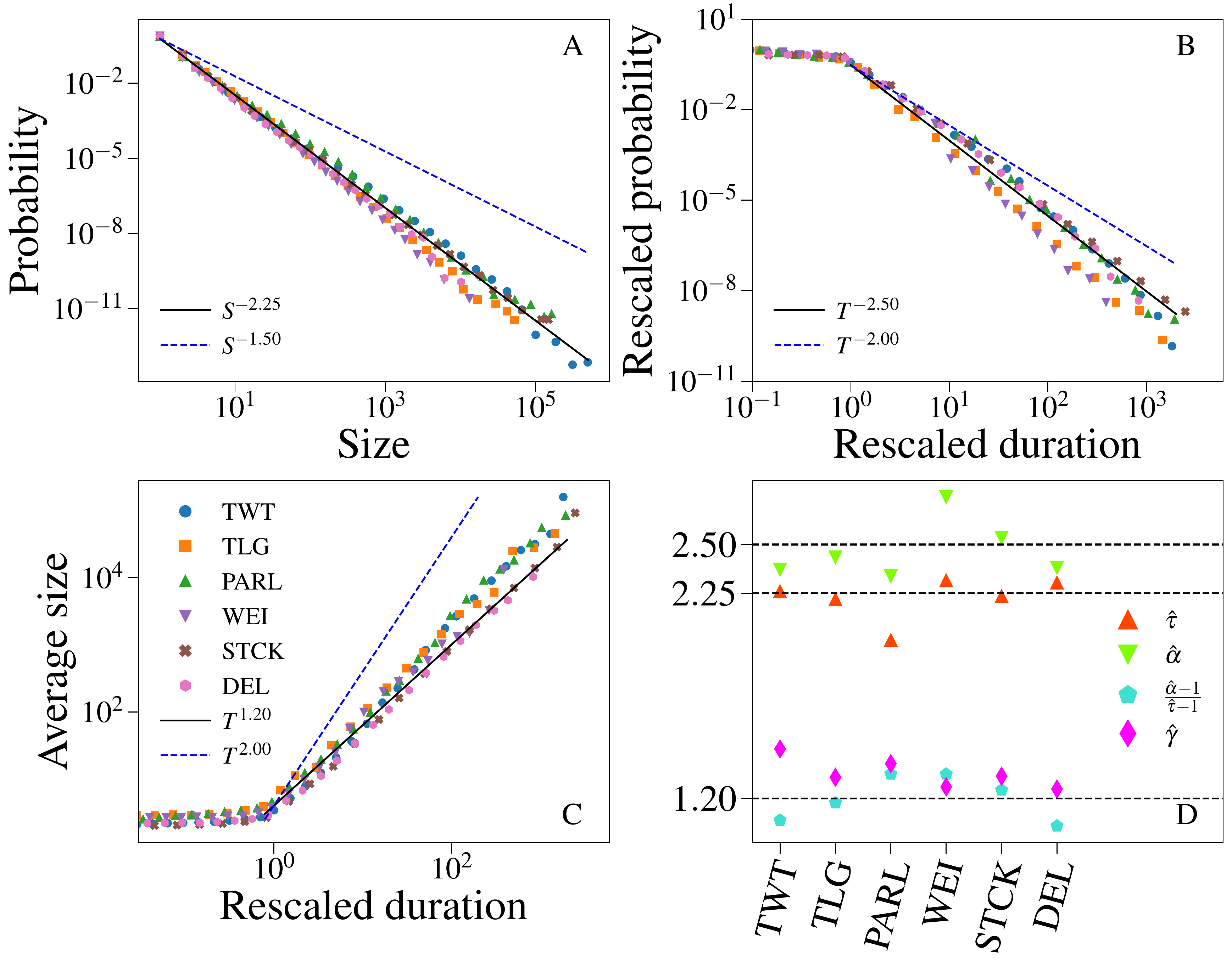}
\end{figure}
\noindent {\bf Fig. 2.} 
{\bf Universality and criticality of information propagation in social media.}
A) Avalanche size distribution. Different colors/symbols indicate data obtained from different social media. Acronyms are defined as in Fig.~1. In this panel, the full line stands for RFIM critical scaling; the dashed line denotes BP critical scaling.
B) Distribution of avalanche duration for the same data as in panel A. To make the distributions collapse one on the top of the other, duration is multiplied by the factor $1/\Delta^*$ and probabilities are multiplied by the factor $\Delta^*$.
C) Average size of avalanches with given duration. Data are the same as in A and B. The abscissa of each curve is rescaled as $\Delta \to \Delta/\Delta^*$.
D) Maximum likelihood estimates of the exponents $\hat{\tau}$, $\hat{\alpha}$
and $\hat{\gamma}$, see SM G for details. We also display the ratio $(\hat{\alpha}-1)/(\hat{\tau}-1)$. Error bars are
always smaller than the size of the symbols. The dashed lines at $\tau = 2.25$, $\alpha = 2.50$ and $\gamma = 1.20$ correspond to best fit with the data (full lines) of panels A, B and C, respectively.

\newpage

\begin{figure}[htp]
    \centering
    \includegraphics[width=0.9\textwidth]{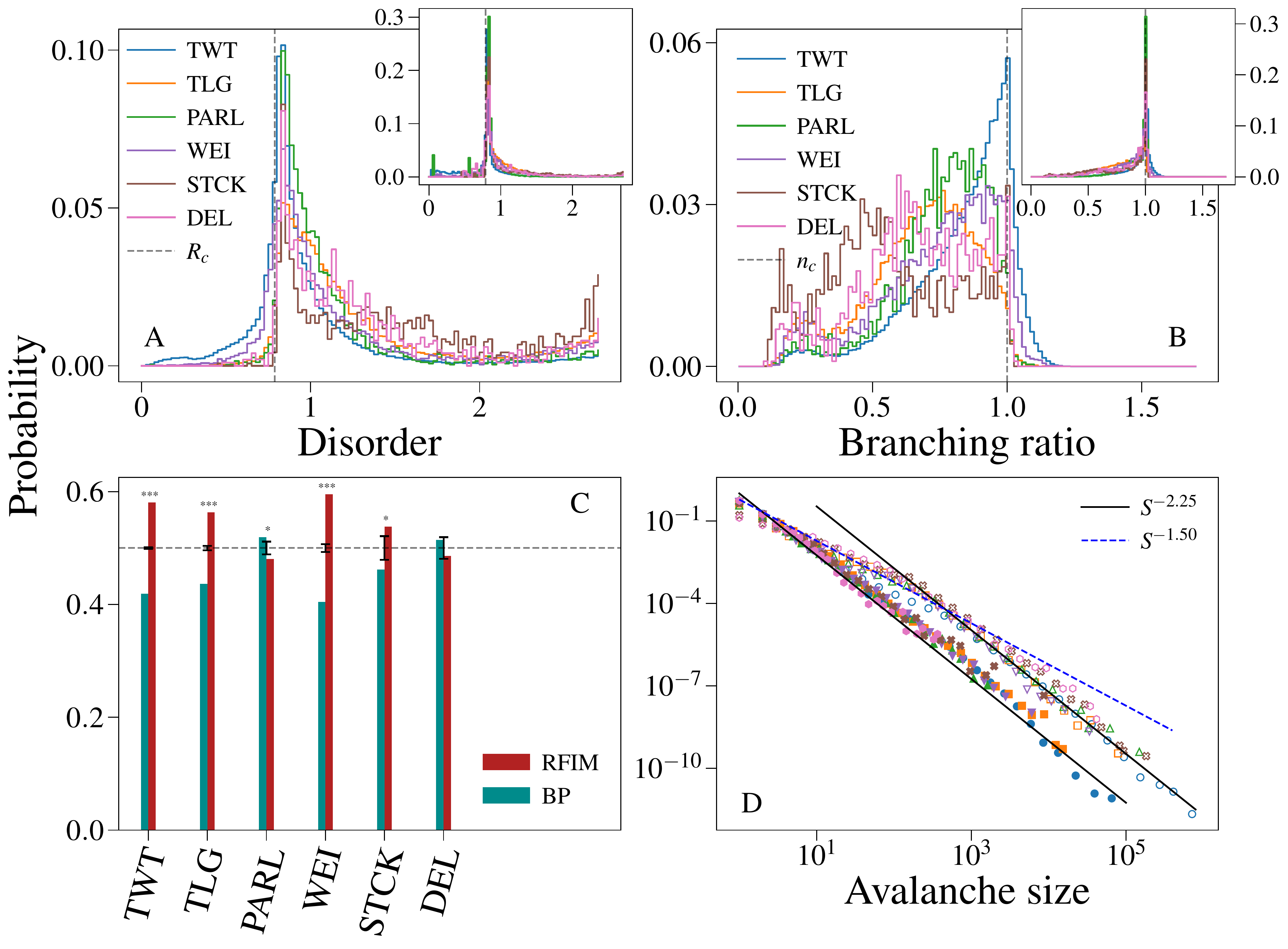}
\end{figure}

\noindent {\bf Fig. 3.} 
{\bf Criticality and complexity of information propagation in online social media.}
A) We fit each individual TS against the RFIM to determine the best estimator of the disorder parameter $\hat{R}$. We then compute the distribution of $\hat{R}$ for all
TS of a given data set. Acronyms of the data sets are defined as in Fig.~1. We fit only avalanches whose size is at least equal to $S_{min}=10$. The dashed vertical grey line denotes $R_c$, i.e., the critical value of the RFIM parameter. The inset shows the same data as in main panel, but each TS contributes to the histogram with a weight equal to its total number of events. B) Same analysis as in A, but obtained by fitting individual TS against the BP model to determine the best estimator  of the branching ratio $\hat{n}$. 
C) Probability that the log-likelihood ratio test favors RFIM over BP (blue), or vice versa BP over RFIM (red). Only TS that are sufficiently well fitted by both models are considered in the analysis, see Fig.~4B. Error bars represent $\sigma/N$, where $N$ is the sample size and  $\sigma = \sqrt{0.25 \, N}$ is the standard deviation of a binomial distribution with probability of success equal to $1/2$. Asterisks are used to denote significant deviations from the unbiased binomial model, i.e., three asterisks indicate  $p<0.001$, and one asterisk stands for $p<0.1$ .
D) We use the classification of panel C to divide TS in two distinct classes. We then consider only TS whose best estimators are sufficiently close to the critical value of the model representing their class, i.e., $|\hat{R}-R_c|/R_c \leq 0.05$ or $|\hat{n}-n_c|/n_c \leq 0.05$, to compute the distribution of avalanche size for each class. Full symbols are used for the RFIM class, empty markers are used to display the distributions of the BP class. Full lines indicate  RFIM critical scaling, while the dashed line denotes BP critical scaling.

\newpage

\newpage

\begin{figure}[htp]
    \centering
    \includegraphics[width=0.8\textwidth]{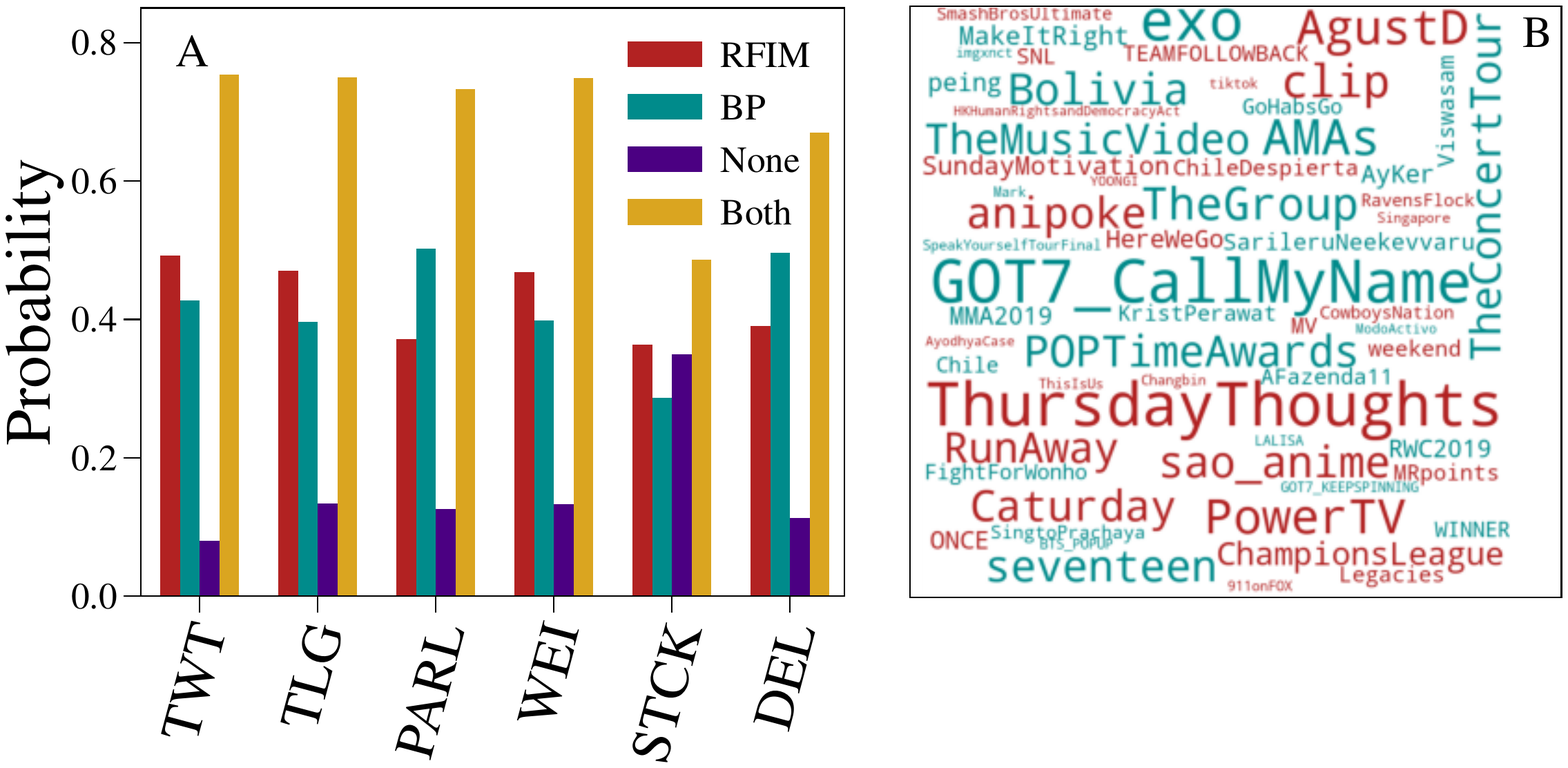}
\end{figure}
\noindent {\bf Fig. 4.} 
{\bf Simple vs. complex contagion in online social media.}
A) We consider avalanches with size $S \geq S_{min}=10$ and fit them against the BP and the RFIM.
For each TS, we establish whether the fits against the individual models 
are statistically significant or not; if both fits can not be rejected, we then 
select the best model by means of the log-likelihood ratio.
We report
the fraction of TS that are classified in the RFIM class. This fact may happen
because the RFIM fit can not be rejected whereas the BP is rejected, or both fits can not be rejected but
RFIM is favored over BP in terms of log-likelihood ratio. 
The fraction of TS that are classified as BP is defined in 
an analogous manner.
The fraction of TS that is classified as neither BP nor RFIM is represented by the bar labeled as 'None.' 
Finally, some TS pass both statistical tests. 
Their fraction is denoted by the label 'Both' in the figure.
In this case, the log-likelihood ratio test is required for model selection, see Fig.~3C.
B) We restrict our attention to Twitter hashtags containing characters from the English alphabet only, and display the $30$ most popular hashtags classified either in the RFIM (blue) or the BP (red) classes. The font size is proportional to the rank of the hashtag in each class. Hashtags of both classes are selected among those that are sufficiently critical, i.e., $|\hat{R}-R_c|/R_c \leq 0.05$ for a TS in the RFIM class or $|\hat{n}-n_c|/n_c \leq 0.05$ for a TS in the BP class.



\newpage
\section*{SUPPLEMENTAL MATERIAL}

\renewcommand{\theequation}{S\arabic{equation}}
\setcounter{equation}{0}
\renewcommand{\thefigure}{S\arabic{figure}}
\setcounter{figure}{0}
\renewcommand{\thetable}{S\arabic{table}}
\setcounter{table}{0}

\subsection{Data sets}


We study data sets concerning the activity of users in six different social media, namely Twitter, Telegram, Parler, Weibo,
StackOverflow and Delicious. 
For each system we identify all (hash)tag in the data and build
a time series (TS) for each (hash)tag. 
The TS contains the 
times, i.e., $\{t_1, t_2,\dots\}$, when
the 
(hash)tag
is observed in the data. 
The Twitter data set is composed of \mbox{2,353,192,777} Tweets corresponding to a $10\%$ random sample of all Tweets posted on Twitter during the observation window Oct. 1 - Nov. 30, 2019. The collection of this data has been performed via the Indiana University OSoME Decahose stream~\cite{osome, twitterdeca}.
Telegram TS
are extracted from a total of \mbox{317,224,715} messages, originally
collected in Ref.~\cite{baumgartner2020pushshift}. Parler TS are extracted from a
total of \mbox{183,062,974} posts, originally collected in
Ref.~\cite{aliapoulios2021early}.
Weibo TS are extracted from \mbox{226,841,249} posts, originally collected in Ref.~\cite{fu2013assessing}. StackOverflow TS are extracted from a 
total number of \mbox{46,947,635} questions and answers. 
Delicious TS were extracted from \mbox{7,034,524} users actions, originally
collected in Ref.~\cite{basile_topical_2015}.
Timestamps always have the temporal resolution of the second, except for the
StackOverflow data set, whose temporal resolution is the millisecond.
Table~1 summarize the properties of these data sets.

\begin{table}[!htb]
\begin{center}
\resizebox{\textwidth}{!}{%
\begin{tabular}{l|r|r|r|r|r}
Data set & Acronym & Temporal window & Time series & Events & URL  \\
\hline \hline
Twitter & TWT & Oct. 1, 2019 - Nov. 30,  2019 & 15,700,708 & 710,124,693 & \cite{TWTdata}   \\
\hline
Telegram & TLG & Sep. 22,  2015 - Jun. 11, 2019 & 5,141,612 & 75,596,578 & \cite{TLGdata} \\
\hline
Parler & PARL & Aug. 1, 2018 - Jan. 11, 2021 & 183,062,974 & 22,831,777 & \cite{PARLdata} \\
\hline
Weibo & WEI & Jan. 2, 2012 - Dec. 30, 2012 & 1,958,768 & 19,560,710 & \cite{WEIdata} \\
\hline
StackOverflow & STCK & Aug. 1,  2008 - Dec. 1, 2019 & 56,525 & 55,084,783 & \cite{SOdata} \\
\hline
Delicious & DEL & Mar. 10, 2007 - Aug. 10, 2011  & 1,052,098 & 21,373,192 & \cite{DELdata} \\
\hline \hline
\end{tabular}
}
\end{center}
\caption{Summary table of the data. From left to right we report: the 
name of the data set, the acronym we use to refer to the data set, 
the temporal window of data collection, the total
number of time series, the total number of events(times) and a link to the original data.
Events correspond to the observation of items in the original
data.}
\end{table}

\subsection{Data cleaning}

\begin{figure}[!htb]
\begin{center}
\includegraphics[width=0.95\textwidth]{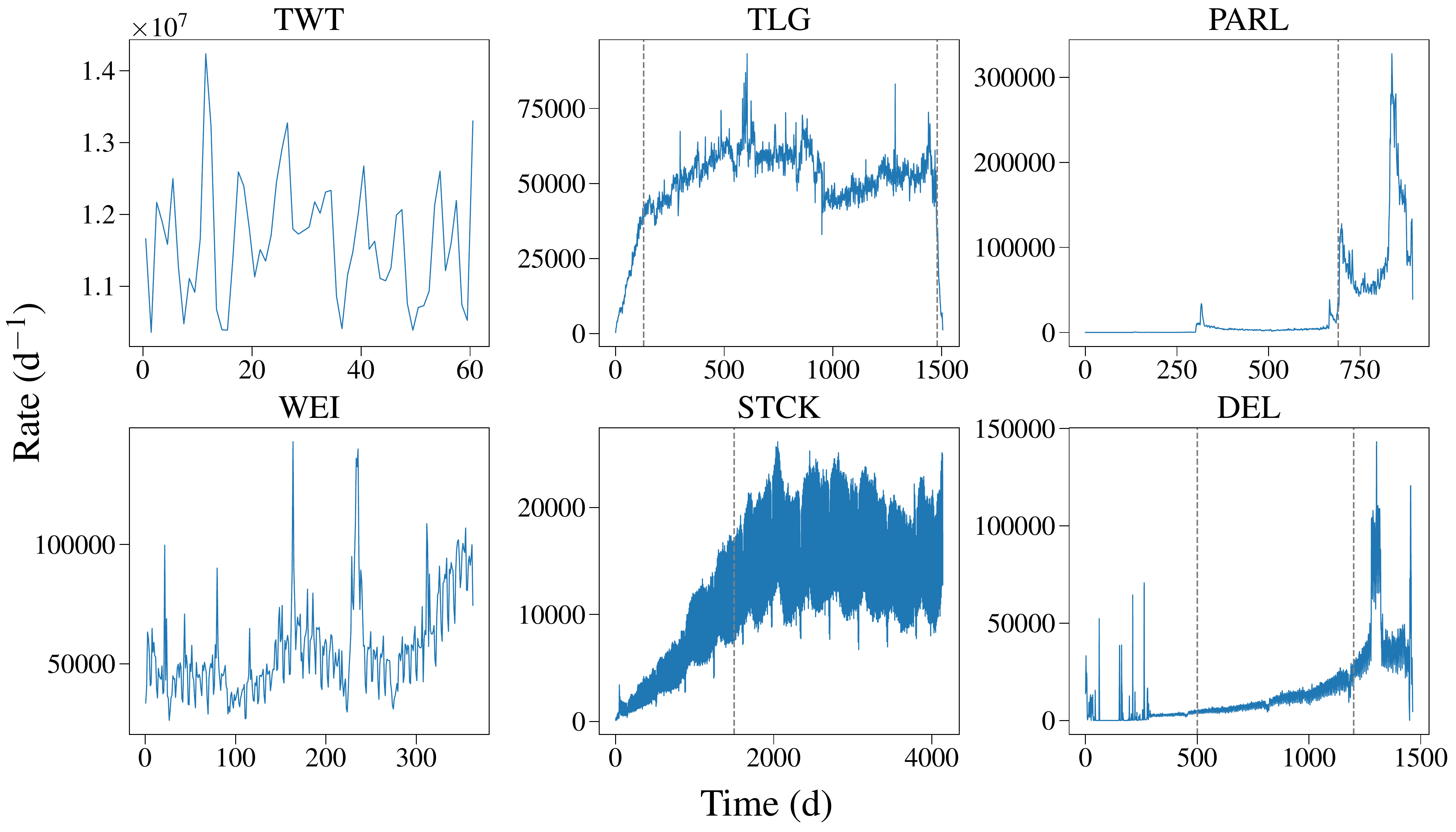}    
\end{center}
\caption{Daily rate of activity in the original data sets. The rate is computed as the
total number of events per day. The dashed vertical lines in the panels for 
TLG and DEL mark
the beginning and the end of the reduced temporal window. 
The dashed vertical line in the panels for PARL and STCK  mark 
the beginning of the reduced
temporal window, which in this case ends where the original window ends.}
\label{fig:rate_init}
\end{figure}

Fig.~\ref{fig:rate_init} shows the daily rate of activity in each data set. While TWT and WEI 
display a rate of activity almost constant over our observation period, the other data sets display significant variations.

We restrict our attention to observation windows where all data are nearly stationary, i.e., the number of events per unit time is roughly constant for time units much larger than the temporal resolution of the data. These shorter observation windows 
are highlighted in Fig.~\ref{fig:rate_init}.

\begin{figure}[!htb]
\begin{center}
\includegraphics[width=0.95\textwidth]{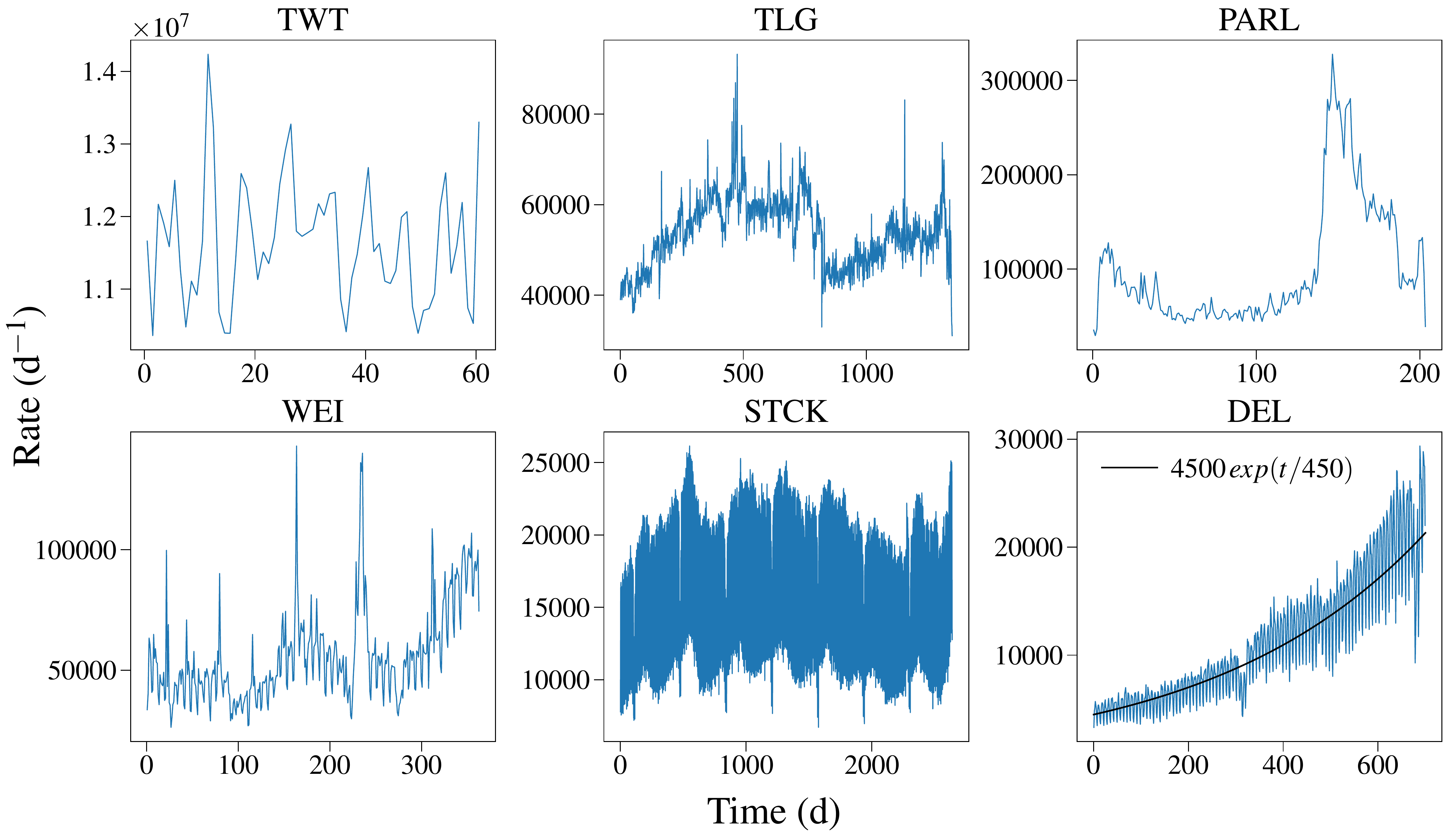}    
\end{center}
\caption{Daily rate of activity in the reduced data sets. The rate is computed as the
total number of events per day. In each panel we show the same data as in
Fig.~\ref{fig:rate_init}, but restricted to the temporal windows
delimited by the dashed vertical lines respectively for each data set.}
\label{fig:rate_final}
\end{figure}

Daily rates in the reduced temporal window are shown separately in 
Fig.~\ref{fig:rate_final}. Table~2 reports information
about the data sets as they result after reducing the temporal windows. 
The results
shown in the main text and in the SM are all obtained from the analysis of data sets over reduced observation windows.

\begin{table}[!htb]
\begin{center}
\resizebox{\textwidth}{!}{%
\begin{tabular}{l|r|r|r|r|r}
Data set & Acronym & Temporal window & Time series & Events & URL \\
\hline \hline
Twitter & TWT &  Oct. 1, 2019 - Nov. 30, 2019 & 15,700,708 & 710,124,693 & \cite{TWTdata}   \\
\hline
Telegram & TLG & 1,350 days & 4,972,879 & 72,593,735 & \cite{TLGdata} \\
\hline
Parler & PARL & 204 days & 753,215 & 20,634,978 & \cite{PARLdata} \\
\hline
Weibo & WEI & Jan. 2, 2012 - Dec. 30, 2012 & 1,958,775 & 20,365,986 & 
\cite{WEIdata} \\
\hline
StackOverflow & STCK & 2,639 days & 55,802 & 45,227,132 & \cite{SOdata} \\
\hline
Delicious & DEL & 700 days & 528,170 & 7,892,075 & \cite{DELdata} \\
\hline \hline
\end{tabular}
}
\end{center}
\caption{Summary table of the data after reduction of the observation windows. 
From left to right we report: the 
name of the data set, the acronym we use to refer to the data set, 
the temporal window of data collection, the total
number of time series, the total number of events (times) and a link to the original data.
}
\end{table}

\subsection{Beyond social media: neuronal systems and earthquakes}

In addition to the six data sets concerning social media,
we further study data sets 
describing activity in different systems.

We consider a set of 88 TS, collected in 
Ref.~\cite{timme2016criticality},
generated by monitoring the spontaneous activity of dissociated cultures
of rat's hippocampal cells. Specifically, we consider the culture number 1 in the 
11-th day in vitro and refer to it as RHDC (Rat Hippocampal Dissociated Cultures).
A set of 166 TS, collected in 
Ref.~\cite{ito2014large,litke2004does}, generated by monitoring the neural 
activity in cultured slices of mice somatosensory cortex is further considered. 
In this case we consider the data set number 1 and refer to it as MSOS 
(Mouse Somatosensory Organotypic Slice). 
We also consider a data set generated by monitoring the neural
activity in the premotor cortex of a macaque, collected in 
Ref.~\cite{lawlor2018linear}. We use the MT\_S2 data set and refer to it as
MPC (Macaque Premotor Cortex). In these systems each electrode
is associated to a TS and an event corresponds to the detection of a spike by the electrode.

\begin{figure}[!htb]
\begin{center}
\includegraphics[width=0.95\textwidth]{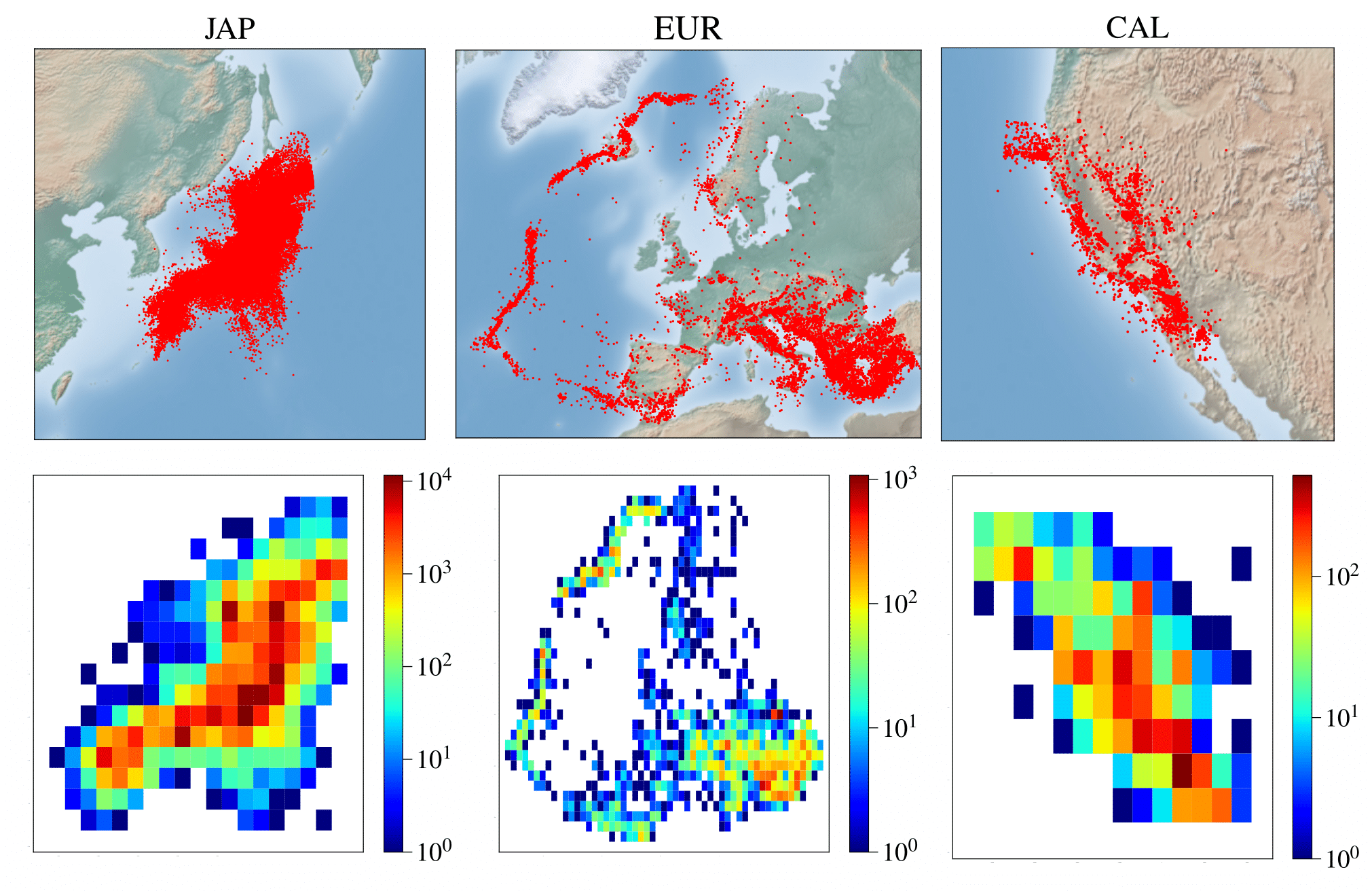}    
\end{center}
\caption{Construction of TS from seismological data. 
From left to right we report: the Japan catalog, the European catalog and the
Californian catalog.
Top row: spatial distribution of earthquakes in the three catalogs considered.
Bottom row: histogram of the spatial distribution. Bins are squares of side
100 Km. }
\label{fig:earthquakes_binning}
\end{figure}

We further consider three catalogues of earthquakes reporting seismological
activity in Japan~\cite{JAPdata}, in California~\cite{CALdata} and in 
Europe~\cite{grunthal2013share}. 
In the case of the California catalogue, we 
discard all events prior to Jan. 1, 1900.
For each of these catalogues, we divide geographical space into bins.  For each bin, we 
construct a TS composed of the time of events whose longitude and latitude falls within the
bin, in the same way as done in Ref.~\cite{karsai2012universal}. 
The procedure of geographical binning is illustrated in 
Fig.~\ref{fig:earthquakes_binning}. Table~3 summarizes the properties of these data sets.

\begin{table}[!htb]
\begin{center}
\begin{tabular}{l|r|r|r|r|r}
Data set & Acronym & Temporal window & Time series & Events & URL \\
\hline \hline
Rat & RHDC & 3,578,396.8 [ms] & 88 & 876,629 & 
\cite{RHDCdata}   \\
\hline 
Mouse & MSOS & 1 [hour] & 166 & 938,018 & 
\cite{MSOSdata}   \\
\hline 
Macaque & MPC & 174,890 [ms] & 46 & 273,244 & 
\cite{MPCdata}   \\
\hline 
Japan & JAP & Jul. 1, 1985 - Dec. 31, 1998 & 192 & 199,446 & 
 \cite{JAPdata} \\
\hline 
California & CAL & Apr. 30, 1900 - Dec. 27, 2000 & 81 & 5,340 & 
 \cite{CALdata} \\
\hline 
Europe & EUR & Jan. 8, 1900 - Dec. 31, 2006 & 638 & 19,126 & 
\cite{EURdata} \\
\hline \hline
\end{tabular}
\end{center}
\caption{Summary table of data sets 
describing neuronal and seismological activity.
From left to right we report: the 
name of the data set, the acronym we use to refer to the data set, 
the temporal window of data collection, the total
number of time series, the total number of events (times) and a link to the original data.
}
\end{table}

\subsection{Defining avalanches from time series}


Given a TS $\{t_1, t_2,\dots\}$,
we define an avalanche starting at $t_b$ as a sequence of events
$\{t_b,t_{b+1},\dots,t_{b+S-1}\}$ such that $t_b - t_{b-1} > \Delta$,
$t_{b+S} - t_{b+S-1} > \Delta$ and $t_{b+i} - t_{b+i-1} \leq \Delta$ for all
$i=1,\dots,S$, where $\Delta$ is the resolution parameter. The size $S$ of
an avalanche is the number of events within it and the duration $T$ is the time lag 
between the first and last event in the avalanche, i.e., $T=t_{b+S-1}-t_b$. 
Depending on the value of $\Delta$, the same TS may correspond to different avalanches.


We follow the principled approach of 
Ref.~\cite{notarmuzi2021percolation}, where avalanches are 
constructed for $\Delta = \Delta^*$. $\Delta^*$ corresponds to the 
critical point of a one-dimensional percolation model that is used to
describe the TS. To this end,
we define the order parameter $P_{\infty}$ of the percolation 
model and its associated
susceptibility $\chi$, respectively, as 
\begin{equation}
    \begin{split}
        & P_{\infty} = \langle S_M \rangle \\
        & \chi = \frac{\langle S_M^2 \rangle - \langle S_M \rangle ^2}
        {\langle S_M \rangle} \, .
    \end{split}
    \label{eq:Delta*}
\end{equation}
Here, $\langle S_M \rangle$ is the average over 
all the 
TS of the size of the largest
avalanche $S_M$. The transition point is associated to the value of 
$\Delta$ where the susceptibility $\chi$ reaches its maximum, 
i.e., $\Delta^* = \text{arg max}\left[ \chi(\Delta) \right]$. 
Note that TS with only
one event introduce an offset in the measure of $P_{\infty}$ and are not 
informative w.r.t. the optimal resolution $\Delta^*$, i.e., $S_M = 1$ for 
any $\Delta$ in these TS. 
For this reason, we remove these TS from the sample and compute $P_{\infty}$
and $\chi$ considering only TS composed of at least two events.

\begin{table}[!htb]
\begin{center}
\begin{tabular}{l||l|l|l|l|l|l r}
Data set & TWT & TLG & PARL & WEI & STCK & DEL \\
\hline 
$\Delta^*$ (s) & 1,566 & 30,549 & 3,845 & 8,413 & 21,135 & 29,853  \\
\end{tabular}
\end{center}
\caption{Summary table of the values of $\Delta^*$ obtained by maximizing the
susceptibility~(\ref{eq:Delta*}) on data sets generated on social media. 
We report the name of the data set (upper row) and
the associated value of $\Delta^*$ (bottom row), expressed in seconds.}
\label{tab:delta_vals1}
\end{table}

Table~\ref{tab:delta_vals1} reports the values of the optimal resolution 
$\Delta^*$ obtained by means 
of the percolation analysis on the social media data sets. 
The avalanche statistics reported in the main text is obtained for $\Delta= \Delta^*$. The statistics refers to all avalanches, excluding the largest one of each TS. This choice is due to the well-known fact that in percolation theory the largest cluster respects a different statistics than that of finite  clusters~\cite{stauffer2018introduction}.

\begin{table}[!htb]
\begin{center}
\begin{tabular}{l||l|l|l|l|l|l r}
Data set & RHDC & MSOS & MPC & JAP & CAL & EUR \\
\hline 
$\Delta^*$ (s) & 4.841 & 10.116 & 1.188 $\cdot 10^{-3}$ & 994,194 & 1,566,860 & 1,678,770 \\
\end{tabular}
\end{center}
\caption{Summary table of the values of $\Delta^*$ obtained by maximizing the susceptibility~(\ref{eq:Delta*}) on data sets not 
representative of social media. 
We report the name of the data set (upper row) and
the associated value of $\Delta^*$ (bottom row), expressed in seconds.}
\label{tab:delta_vals2}
\end{table}

Table~\ref{tab:delta_vals2} reports the values of the optimal resolution 
$\Delta^*$ for the data sets not concerning activity in social media.

\subsection{The branching process}


We consider an homogeneous mean-field branching process (BP), where an individual initially active spreads activity to a random number of peers, who can in turn spread activity further.
The process continues for a number $T$ of time steps or generations, until there is a generation
in which no individual further spreads activity. $T$ is the duration of the avalanche. The size $S$ of the avalanche is given by the total number of individuals activated during the avalanche.
The only tunable parameter of the model is denoted by $n$, representing
the average number
of individuals who 
are activated
from a single spreader. $n$ is known as the branching
ratio. BP  is critical for $n=n_c=1$. 

Finite avalanches of activity in the BP obey the laws
\begin{equation}
    \begin{split}
        & P(S) = S^{-\tau} \mathcal{D}_S(S^{\sigma} n')\\
        & P(T) = T^{-\alpha}  \mathcal{D}_T(T^{1/z \nu} n')\\
        & \langle S \rangle(T) \propto  T^{\gamma} \, ,
    \end{split}
    \label{eq:exponents}
\end{equation}
where $\langle \cdot \rangle$ is the average over different avalanches, and 
$P(S)$ and $P(T)$ are the probability distributions of $S$ and $T$, respectively.
The functions $\mathcal{D}_S$ and $\mathcal{D}_T$ are known as scaling functions
and introduce a correction at small values of their argument, where  we have
defined the reduced distance from the critical point $n' = |n-n_c|/n_c$.
The above exponents are not independent, rather they are related 
by $\gamma = 1/(\sigma z \nu) = (\alpha-1)/(\tau-1)$.
For BP, we have that $\tau = 3/2$ and $\alpha=2$. $\sigma$, $z$ and $\nu$ are additional critical exponents. We do not explicitly consider them in our analysis.

\subsection{The Random Field Ising Model}

We consider
the mean-field formulation of the zero-temperature 
Random Field Ising Model (RFIM).
In the RFIM, agent $i$ is characterized by
the state variable $y_i = \pm 1$ 
indicating whether the agent is active, i.e.,
$y_i = + 1$, or not, 
i.e., $y_i = -1$.
In the initial configuration all agents are inactive.
In the long-term limit, all agents become active. Activation of individual agents may happen at very different stages of the dynamics.  However, once in the active state, agents can not change their state back to inactive.
Each agent $i$ has a propensity $h_i$ to become active, with $h_i \in (-\infty, +\infty)$.
A large value of 
$h_i$ indicates that the agent is particularly prone to 
become active. 
Agents interact by means of ferromagnetic interactions that model social pressure,
i.e.,
active neighbors push an inactive agent to become active.
The whole system is further affected by public information which 
all agents have access to and that pushes users toward 
becoming active
with intensity $H \in (-\infty, +\infty)$. 

In the initial configuration, all agents are inactive
($y_i=-1$ for all agents $i$). External pressure $H$ grows till 
the agent with the largest $h_i$ value becomes active.
This change of state can trigger
an avalanche of activity in the other nodes.
Specifically, agent $j$ becomes active if the following condition is met
\begin{equation}
H + h_j + N^{-1} \sum_{k \neq j} y_k > 0 \, ,
\label{eq:threshold}    
\end{equation}
where $N$ is the system size and the mean-field formulation is expressed by the all-to-all
interaction.
When an avalanche ends, the external pressure $H$ grows again until 
a new user becomes active and triggers a new avalanche.
The field is frozen during the unfolding of avalanches, meaning that
avalanches are characterized by a time scale much shorter than the one characterizing external pressure.
The size $S$ of an avalanche is given by the number of users 
that are activated during the avalanche;
its duration $T$ is given by 
the activation rounds characterizing  the avalanche. 
The stochasticity of the model comes from the random nature of the propensities
$h_i$, extracted from a normal distribution with zero mean and variance $R$. 
The choice of the normal distribution is quite standard both for ferromagnets
and for social systems~\cite{michard2005theory}.
$R$ is the control parameter of the model, which is critical for $R=R_c=\sqrt{2/\pi}$.
Avalanche statistics still obey laws similar to those of Eqs.~(\ref{eq:exponents}). The functional form of the scaling functions differs from those of BP; also, their argument is given in terms of the distance from the critical point of RFIM, i.e., $n' = |n-n_c|/n_c$ is replaced by $R' = |R-R_c|/R_c$. The values of the critical exponents are $\tau = 9/4$ and $\alpha = 7/2$~\cite{sethna2001crackling}.

\subsection{Estimation of the exponents}


To estimate the exponents $\hat{\tau}$ and $\hat{\alpha}$ 
for the empirical avalanche distributions, we use the fact that
for a generic power-law probability distribution with exponent $\eta$, 
the maximum likelihood estimator can be written as
\begin{equation}
    \hat{\eta} = 1 + Z \left( \sum_{i=1}^Z \log \frac{x_i}{x_{min}} \right) \, ,
    \label{eq:MLE}
\end{equation}
where $x_i$ is a data point of the empirical sample, and
$x_{min}$ is the smallest value of the 
sample
that is expected to truly 
respect the power-law statistics~\cite{clauset2009power}. $Z$ is the number of data points $x_i \geq x_{min}$.
If the variable under consideration is discrete,
the factor $x_{min}$ in the denominator of the 
logarithm in Eq.~(\ref{eq:MLE}) must be replaced by
$x_{min}-0.5$. 
The error on the maximum likelihood estimator is 
$\Delta \hat{\eta} = (\hat{\eta}-1)/\sqrt{Z}$. 
We use $S_{min}=2$ to fit the size distribution and $T_{min}=2\Delta^*$ to 
fit the duration distribution. This protocol
allows us to measure $\hat{\tau}$ and $\hat{\alpha}$ of the distributions $P(S)$ 
and $P(T)$, respectively, and to further measure the scaling exponent
$\hat{\gamma}$ as $(\hat{\alpha}-1)/(\hat{\tau}-1)$.
Assuming that the two estimators are
uncorrelated, the uncertainty on the
ratio $f(\hat{\tau}, \hat{\alpha}) = (\hat{\alpha}-1)/(\hat{\tau}-1)$ can be simply evaluated as $\sqrt{\left(\frac{\partial f}{\partial \hat{\tau}} \Delta \hat{\tau}\right)^2 + \left(\frac{\partial f}{\partial \hat{\alpha}} \Delta \hat{\alpha}\right)^2 }$.


To independently estimate the exponent $\hat{\gamma}$,
we take the logarithm 
of both sides in the last Eq. in~(\ref{eq:exponents})
and perform linear regression. The exponent $\gamma$ and its uncertainty 
are then given by 
$\hat{\gamma} = \frac{\sum_i (X_i-\langle X \rangle)(Y_i-\langle Y \rangle)}{\sum_i (X_i-\langle X \rangle)^2} $ and $\Delta \hat{\gamma} = \sqrt{\frac{\frac{1}{Z-2} \sum_i \epsilon_i^2}{\sum_i(X_i-\langle X \rangle)^2}}$ respectively, where $X = \log T$ and $Y = \log \langle S \rangle$,
and $\epsilon$ are the residuals.

\subsection{Scaling in neuronal systems and earthquakes}

\begin{figure}[!htb]
\begin{center}
\includegraphics[width=0.95\textwidth]{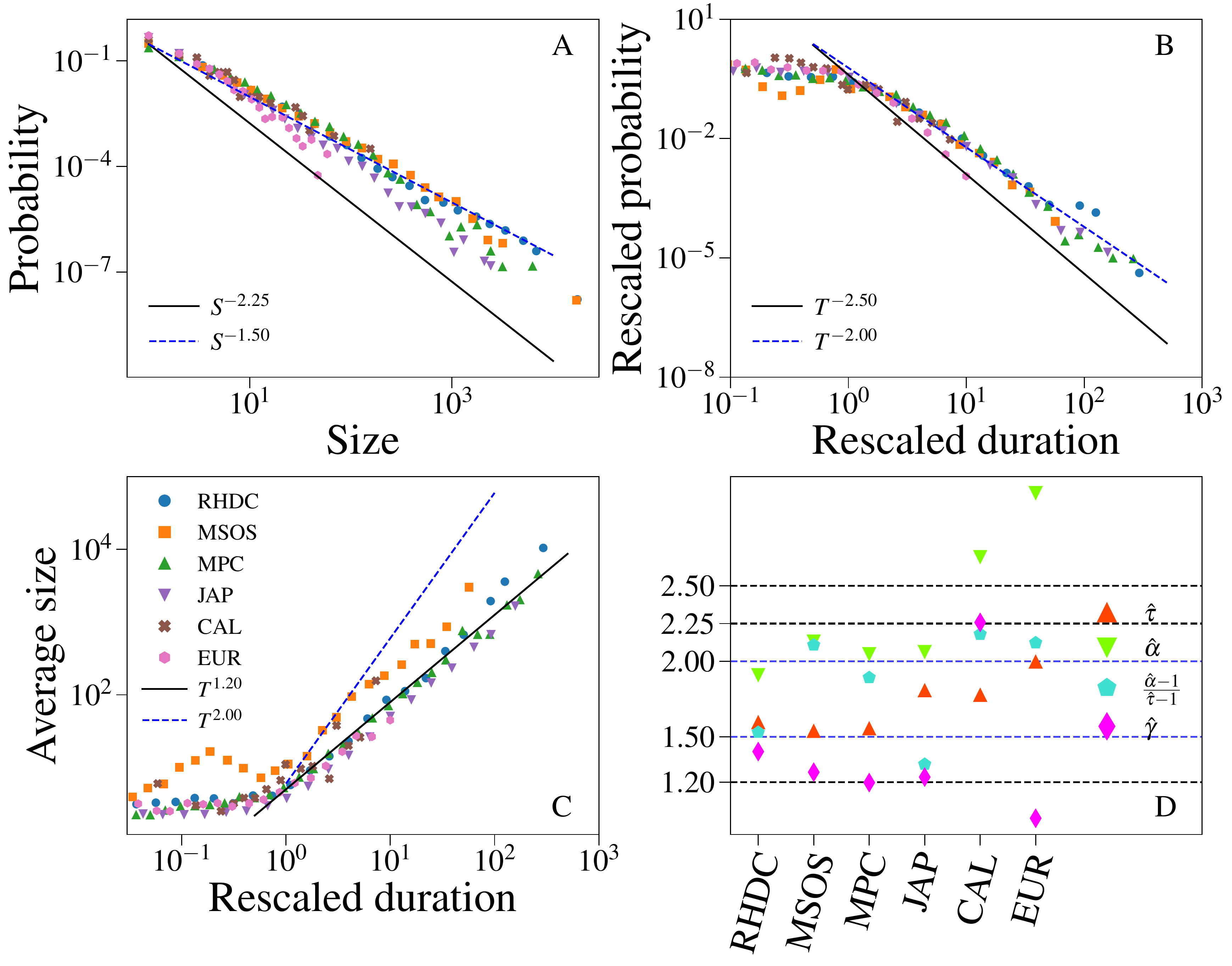}    
\end{center}
\caption{Avalanche statistics in systems other than social media.
A) Distribution of avalanche size. 
Different colors/symbols refer to different systems: rat's hippocampal dissociated cultures (RHDC), mouse somatosensory organotypic slices (MSOS), macaque premotor cortex (MPC), 
earthquakes in Japan (JAP), California (CAL) and Europe (EUR).
The full line stands for RFIM critical scaling; the dashed line denotes BP critical scaling.
B) Distribution of avalanche duration for the same data as in panel A. Duration is rescaled by the factor $1/\Delta^*$ and
probabilities are rescaled by the factor $\Delta^*$. 
The dashed line denotes BP critical scaling.
C) Average size of avalanches with given duration. Data are the same as in A and B. The abscissa of each curve is rescaled by $1/\Delta^*$. 
The dashed line denotes BP critical scaling. 
D) Maximum likelihood estimates of the exponents $\hat{\tau}$, $\hat{\alpha}$
and $\hat{\gamma}$, see SM G for details. }
\label{fig:OtherSysts_avalanches}
\end{figure}

We perform on the supplementary data sets
the same analysis performed in the main text for data sets concerning social media. Results are shown in
Fig.~\ref{fig:OtherSysts_avalanches}. 
The three data sets 
describing neuronal brain activity
in different animals all display the BP statistics for both the size and the duration distributions. The finding is consistent with previous studies~\cite{beggs2003neuronal,friedman2012universal,haldeman2005critical}.
The scaling relation between $\langle S \rangle$ and $T$ does not show the
scaling $\gamma=2$ as expected from BP theory. However, a slightly 
superlinear relation between these quantities has been reported for many 
different neuronal 
systems~\cite{friedman2012universal,fontenele2019criticality}. 

\subsection{Temporal resolution and avalanche statistics}

\begin{figure}[!htb]
\begin{center}
\includegraphics[width=0.95\textwidth]{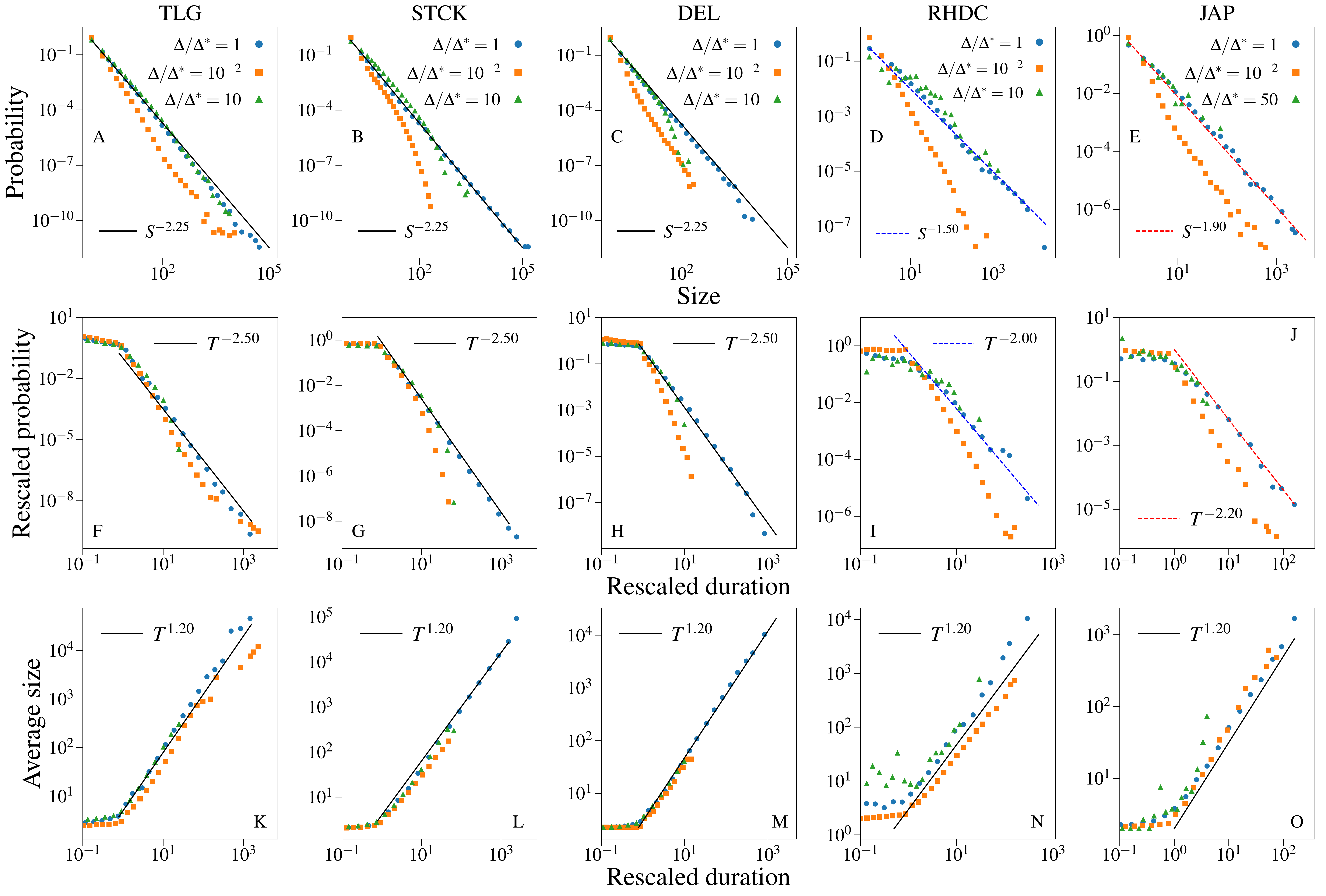}    
\end{center}
\caption{Statistics of avalanches obtained using different values of the temporal resolution.
From left to right we show Telegram (TLG), Stackoverflow (STCK), Delicious (DEL),
rat's hippocampal dissociated cultures (RHDC) and
earthquakes in Japan (JAP). From top to bottom we show the avalanche size distribution,
the avalanche duration distribution and the average size of avalanches with given 
duration. The abscissa in the second and third rows is rescaled by the factor $1/\Delta$
and the ordinate in the second row is rescaled by the factor $\Delta$. Solid black
lines represent the scaling reported for social media in the main text, i.e.,
$\tau=2.25, \alpha=2.5, \gamma=1.2$, the dashed blue line represent the BP scaling,
i.e., $\tau=1.5, \alpha=2$, and the dashed red lines represent the scaling 
$\tau=1.9$ and $\alpha=2.2$.
}
\label{fig:NonOptimal_Delta}
\end{figure}

\begin{figure}[!htb]
\begin{center}
\includegraphics[width=0.95\textwidth]{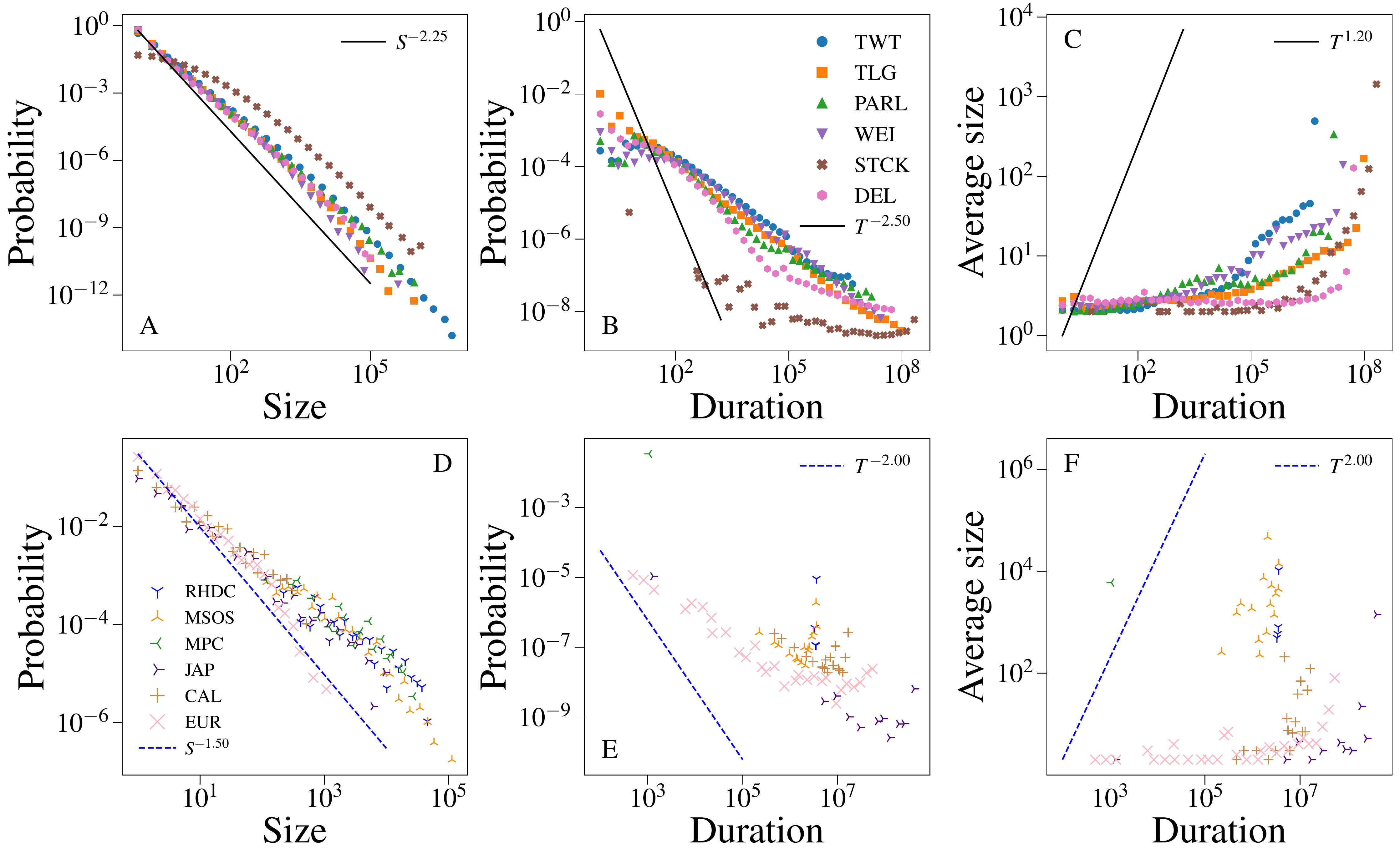}    
\end{center}
\caption{Avalanche statistics assuming 
each TS is an avalanche. 
A) Avalanche size distribution in social media. The solid black line is
the critical RFIM scaling $\tau=2.25$.
B) Avalanche duration distribution for the same data as A. The solid black line is
the scaling $\alpha=2.5$ reported in the main text.
C) Average size of avalanches with given duration for the same data as A. 
The solid black line is the scaling $\gamma=1.2$ reported in the main text.
D) Avalanche size distribution in systems other than social media. The dashed blue 
line is the critical BP scaling $\tau=1.5$.
E) Avalanche duration distribution for the same data as D. The dashed blue line is
the critical BP scaling $\alpha=2$.
F) Average size of avalanches with given duration for the same data as D. 
The dashed blue line is the critical BP scaling $\gamma=2$.
}
\label{fig:DeltaInfty}
\end{figure}

In Fig.~\ref{fig:NonOptimal_Delta}, we display the avalanche statistics of  different systems obtained for
different values of the temporal resolution $\Delta$. For $\Delta \neq \Delta^*$, the  power-law scaling is affected by apparent exponential cutoffs.
The finding is
in perfect agreement with theoretical arguments~\cite{notarmuzi2021percolation}. 
As the avalanche statistics obtained with the present approach represents the correlations
existing in the system~\cite{karsai2012universal}, the observation of distorted distributions means
that the correlations existing in the data have not been properly identified.
The same issue arises when each TS is assumed to be a unique avalanche, see Fig.~\ref{fig:DeltaInfty}. 

\subsection{The scaling function in the distribution of avalanche durations}

\begin{figure}[!htb]
\begin{center}
\includegraphics[width=0.95\textwidth]{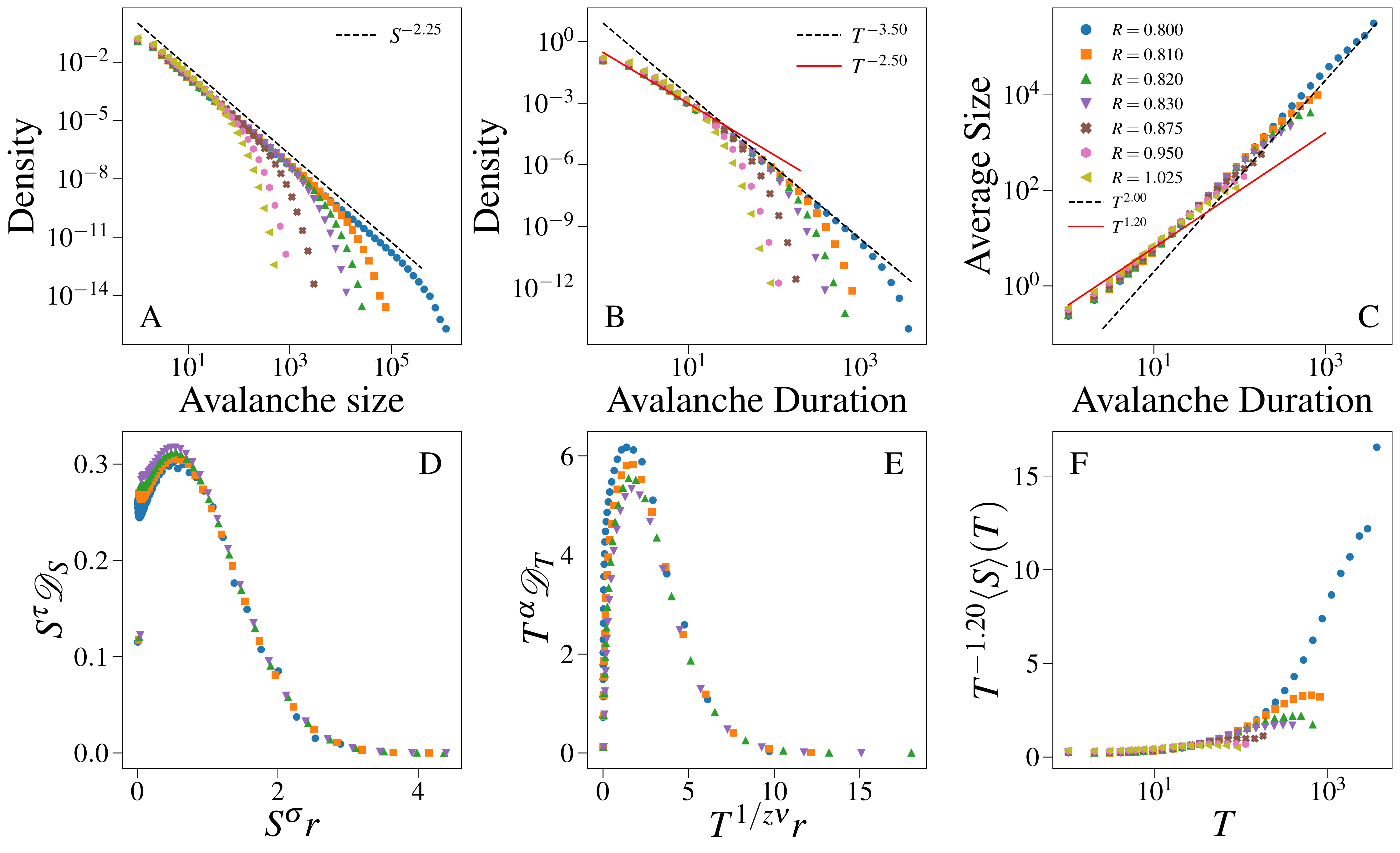}    
\end{center}
\caption{Avalanche statistics in the RFIM on a complete graph 
of size $N=10^9$.
A) Avalanche size distribution. The dashed black line scales as $S^{-\tau}$ with
$\tau=2.25$.
B) Avalanche duration distribution. The dashed black line scales as $T^{-\alpha}$ 
with $\alpha=3.5$, while the solid red line scales as $T^{-2.50}$ as 
our data do. 
C) Average size of avalanches with given duration. The dashed black line scales as $T^{\gamma}$ 
with $\gamma=2$, while the solid red line scales as $T^{1.20}$ as 
our data do. 
D) Scaling function of the size distribution.
E) Scaling function of the duration distribution.
F) Average size of avalanches with given duration, rescaled by $T^{-1.20}$.
}
\label{fig:RFIM}
\end{figure}

The scaling function $\mathcal{D}_T$ appearing in the second Eq. of~(\ref{eq:exponents})
quickly goes to 
a constant value that is independent of $T$
in the limit of large values of its argument, so
that $P(T)$ shows the 
pure power-law decay $T^{-\alpha}$ in such a regime.
The scaling function, however,
introduces a correction 
to the pure power-law scaling
at small values of its argument. As stated in the
main text, the correction is rather strong for the RFIM in large dimension.
The same phenomenology is experienced by the  distribution 
of avalanche sizes $P(S)$, but
the correction is much smaller in this case.
Fig.~\ref{fig:RFIM} shows that the correction on $P(S)$ is nearly 3, while the
correction on $P(T)$ is larger than 50, the correction being computed as 
the ratio between the maximum of the scaling function and its value in the
limit of small argument.
Note that the correction on $P(S)$
in dimension 3 is about 10,
and this  is already sufficient 
to lead to an inaccurate estimation of the asymptotic exponent value~\cite{sethna2001crackling}.
Fig.~\ref{fig:RFIM} C and F further shows how this correction affects the measure of $\gamma$.

\subsection{Models fitting and comparison}
To determine if each single TS can be ascribed to one, both or none of the models we apply the following procedure:
\begin{enumerate}

    \item We fit separately the two models via numerical simulations and likelihood maximization.
    
    \item We evaluate the $p$-value of the fits.
    
    \item We compare the $p$-values of both models and assign the empirical TS to one of three classes: RFIM, BP, none.
\end{enumerate}
\paragraph{Fit.}Given a TS composed of $N$ events, we first compute the probability distribution $P(S)$ of the avalanche sizes identified in the TS.
By means of numerical simulations,
we then compute the conditional distributions of the avalanche size
$Q_{RFIM}(S|R)$ and $Q_{BP}(S|n)$
obtained respectively for the RFIM and BP for a given value of the 
parameters $R$ and $n$ (for exactly the same number $N$ of events).

We note that
the construction of the model distributions $Q$ requires a discretization of 
the parameter space of the models. $R$ varies in the interval $[0.025, 2.7]$ by steps of length  $dR=0.025$ for the RFIM (108 intervals in total).
For the BP, $n$ varies in $[0.02, 1.7]$ by steps of length $dn=0.015$ (113 total intervals).
$dR$ ($dn$) represents the uncertainty on the RFIM (BP) 
parameter. Instead of sampling avalanches from the model at a precisely given value of $R$ ($n$), we consider model instances corresponding to $R$ ($n$) values uniformly distributed  over an interval of length $dR$ ($dn$) centered at $R$ ($n$). Fitting a TS to a model with a specific 
parameter value means estimating the best parameter with an accuracy of
$dR$ ($dn$) for the RFIM (BP). 
The distribution $Q$ corresponding to a specific value of the parameter model is 
constructed as the superposition of $500$ distributions whose parameter values are randomly sampled from the corresponding interval.

Given the empirical distribution $P$ and the model distribution $Q$, we evaluate the log-likelihood function
\begin{equation}
    \text{L}(P || Q) = \sum_{S \geq S_{min}} P(S)\log[Q(S)]  \, .
    \label{eq:KL}
\end{equation}
The summation is performed over all avalanches with $S \geq S_{min}$,
a parameter we vary in our analysis. The distributions $P$ and $Q$ are normalized over the interval $[S_{min}, \infty)$ to account for this fact.
The best fit of the empirical distribution obtained from a TS against the model at hand is obtained by finding the parameter value that maximizes the log-likelihood of Eq.~(\ref{eq:KL}). To avoid numerical problems in the estimation of the likelihood, we smoothen the function $Q$. Details are provided in SM L.

\paragraph{P-value.}To assign a $p$-value to a fit, we follow the prescription of 
Ref.~\cite{clauset2009power}. Let us indicate with $Z_{tail}/Z$  the fraction
of avalanches with $S \geq S_{min}$ in the fitted TS. A synthetic sample of $Z$ avalanches is
created by sampling avalanches with $S \geq S_{min}$
from the selected model with probability $Z_{tail}/Z$ and by sampling avalanches with $S < S_{min}$ 
from the empirical distribution with 
complementary probability. Each of these synthetic samples 
is fitted analogously to the 
original sample obtained from the TS. 
Once a distribution $Q$ is selected 
by means of likelihood maximization, the Kolmogorov-Smirnov (KS) distance is
computed, for both the original sample and
the synthetic sample. The $p$-value of the fit is defined as the fraction of 
synthetic samples whose KS distance from the selected model is larger than the KS distance
between the real sample and its best model. 
\paragraph{Model comparison.}The hypothesis that the real sample has been generated by a
certain dynamical model, say RFIM, 
can not be rejected if 
the $p$-value of the fit to the RFIM is larger than a pre-established significance threshold. We set the threshold to 0.1 in the main text, following the prescription of Ref.~\cite{clauset2009power}.  
Tests of robustness against the choice of this parameter value are reported in SM O.

If one of the two hypotheses can be rejected 
but the other can not be rejected, the non-rejected model automatically becomes the selected model.
If both hypotheses can be rejected, the TS is classified as ``None.''
If,  however, both hypotheses can not be rejected, 
we select as the best model the one with the largest likelihood~\cite{clauset2009power}. We neglect the possibility that a single TS could be a mixture of models.

Empirical data are fitted only if the TS contains at least 50 events
and at least 10 avalanches. 
Technical details of how to efficiently compute the KS distance are given in SM M. We also validate our method on synthetic data and show the robustness of our results
against variations of some parameter values in SM N, O respectively.

\subsection{Calculation of the likelihood}

\begin{figure}[!htb]
\begin{center}
\includegraphics[width=0.95\textwidth]{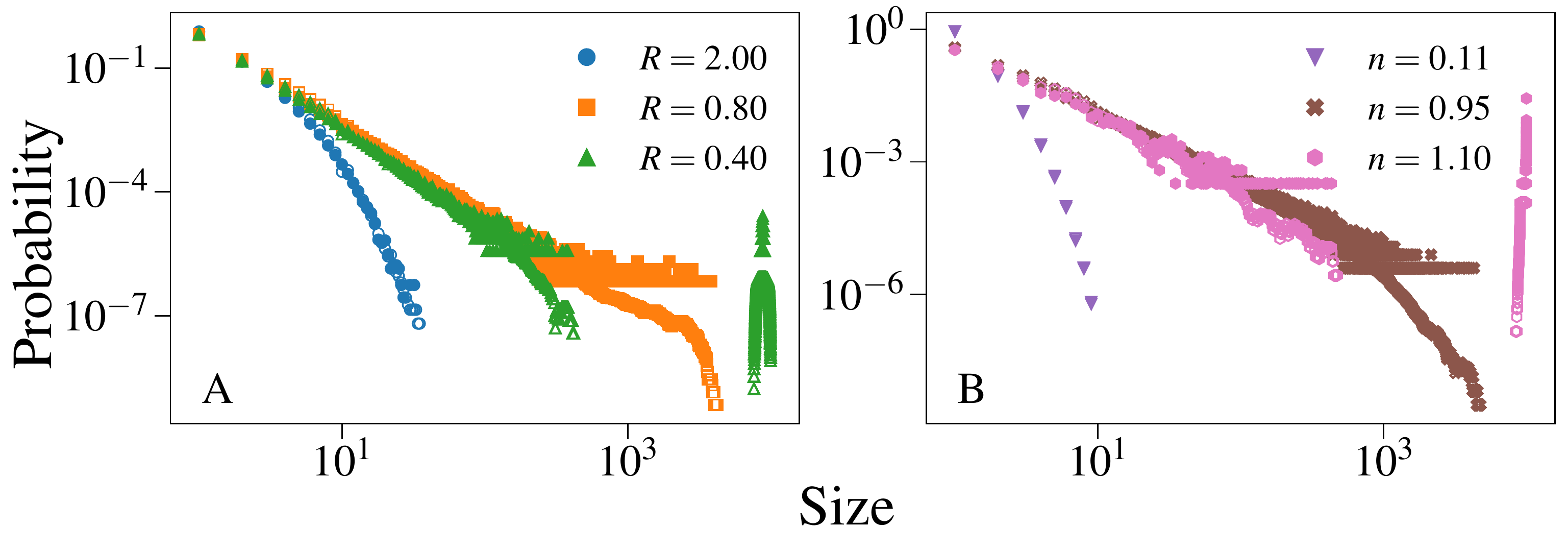}    
\end{center}
\caption{Avalanche size distributions before and after smoothing. In all panels, full
symbols represent the original distribution and empty symbols represent the distribution
after smoothing. We use here $h=0.1$ and set the system size to be $N=10^4$. 
A) Three configurations of the RFIM.
B)
Three configurations of the BP. Each panel shows a subcritical, a nearly
critical and a supercritical configuration.}
\label{fig:smoothing}
\end{figure}

The model distribution $Q$ is estimated from numerical simulations. As such,
finite-size distortions may be present in the tail of the empirical distribution, potentially leading to mistakes in the maximum likelihood fit. 
We therefore apply a rectangular kernel  to 
regularize
the empirical distribution. The width of the rectangular kernel grows exponentially with respect to its argument value at rate $h$.
In Fig.~\ref{fig:smoothing}, we show
the distributions $Q$ before and after smoothing for 
several configurations of both RFIM And BP. The analysis is performed by setting  $h=0.1$. 
This is the value of the smoothing parameter used to obtain the results in  the main text. We verified that results are robust against small variations of $h$,
e.g., $h=0.2$ or $h=0.05$. Note that 
a small variation of $h$ is a significant variation for the width of
the rectangular kernel. 

Once the smoothing is performed, the smoothed distributions represent the 
theoretical model. As such, we use the smoothed distributions for
the calculation of the likelihood, for the calculation of the KS distance,
and for the generation of the synthetic samples required to estimate the $p$-value.

\subsection{Efficient computation of the KS distance}

Let us indicate with $C_P$ and $C_Q$ 
 the Cumulative Distribution Functions (CDFs) of the distributions
$P$ and $Q$, respectively. 
The Kolmogorov-Smirnov (KS) distance between
$P$ and $Q$ is defined as
\begin{equation}
    \text{KS}(C_P, C_Q) = \text{max}_y |C_P(y) - C_Q(y)| \, .
    \label{eq:KS}
\end{equation}

To speed up the computation of Eq.~(\ref{eq:KS}), we rewrite it as
\begin{equation}
    KS(C_P, C_Q) = 
   \text{max}_{i=0,\dots,Z} \{|C_P(S_i) - C_Q(S_i)|, 
    |C_P(S_{i+1}-1) - C_Q(S_{i+1}-1)|\} \, .
    \label{eq:KS_true_vals}
\end{equation}
In the above equation, $1\leq S_1 < S_2 < . . . S_i < S_{i+1} < . . . < S_Z \leq N$ are the sizes of the avalanches used to construct the empirical distribution $P$. By definition of CDF, we have that
$C_P(S_i) < C_P(S_{i+1})$ 
for all
$i=0,...,Z$, where we relied on the conventions $S_0 = -\infty$ and $S_{Z+1}>S_Z$, thus 
$C_P(S_0)=0$ and $C_P(S_{Z+1})=1$.

Estimating the KS distance via Eq.~(\ref{eq:KS_true_vals}) requires to compute the difference between  $C_P$ and $C_Q$ for a number of values of their arguments that is (much) smaller than the one required by the straight implementation of Eq.~(\ref{eq:KS}), since $\{S_i\}$ is a subset of $\{y\}$ containing only the observed values of $S$.

To prove
that Eq.~(\ref{eq:KS_true_vals}) holds we need to show that, for each
$i = 0, \ldots, Z$, we have that
\begin{equation}
    \text{max}_{y \in [S_i,S_{i+1}-1]} |C_P(y) - C_Q(y)| = 
    \text{max} \{|C_P(S_i) - C_Q(S_i)|, 
    |C_P(S_{i+1}-1) - C_Q(S_{i+1}-1)|\} \, .
    \label{eq:aa}
\end{equation}
The validity of the above equation follows from the facts 
that both $C_P$ and $C_Q$ are non-decreasing functions, and
that $C_P$ is constant in the interval $[S_i, S_{i+1}-1]$. As a matter of fact,
 Eq.~(\ref{eq:aa}) is representative for the only three possible cases that can happen:
\begin{enumerate}
    \item $C_P(S_i) \leq C_Q(y)$ for each  $y\in[S_i, S_{i+1}-1]$. Then,
    for each $y$ in this interval,
    \begin{equation}
        |C_P(y) - C_Q(y)| = C_Q(y) - C_P(S_{i+1}-1) \leq C_Q(S_{i+1}-1) - C_P(S_{i+1}-1) \, ,
    \end{equation}
    so that
    \begin{equation}
        \text{max}_{y \in [S_i,S_{i+1}-1]} 
        |C_P(y) - C_Q(y)| = 
        |C_P(S_{i+1}-1) - C_Q(S_{i+1}-1)| \, .
        \label{eq:KS_case1}
    \end{equation}
    \item $C_P(S_i) \geq C_Q(y)$ for each  $y\in[S_i, S_{i+1}-1]$. Then,
    for each $y$ in this interval,
    \begin{equation}
        |C_P(y) - C_Q(y)| = C_P(S_i) - C_Q(y) \leq C_P(S_i) -  C_Q(S_i)\, ,
    \end{equation}
    so that
    \begin{equation}
        \text{max}_{y \in [S_i,S_{i+1}-1]} 
        |C_P(y) - C_Q(y)| = 
        |C_P(S_{i}) - C_Q(S_{i})| \, .
        \label{eq:KS_case2}
    \end{equation}
    \item $\exists \, y^* \in [S_i, S_{i+1}-1]$ such that $C_P(S_i) \geq C_Q(y)$ for 
    $y \in [S_i, y^*]$ and $C_P(S_i) \leq C_Q(y)$ for $y \in [y^*, S_{i+1}-1]$. 
    In this case, the interval $[S_i, y^*]$ can 
    be treated as the former case 2 while the interval $[y^*, S_{i+1}-1]$ can be treated
    as the former case 1, so that in the present case 3 we have
    \begin{equation}
        \text{max}_{y \in [S_i,S_{i+1}-1]} 
        |C_P(y) - C_Q(y)| = 
        \text{max} \{ |C_P(S_{i}) - C_Q(S_{i})|,
        |C_P(S_{i+1}-1) - C_Q(S_{i+1}-1)| \} \, .
        \label{eq:KS_case3}
    \end{equation}
\end{enumerate}

\subsection{Validation on synthetic samples}

\begin{figure}[!htb]
\begin{center}
\includegraphics[width=0.95\textwidth]{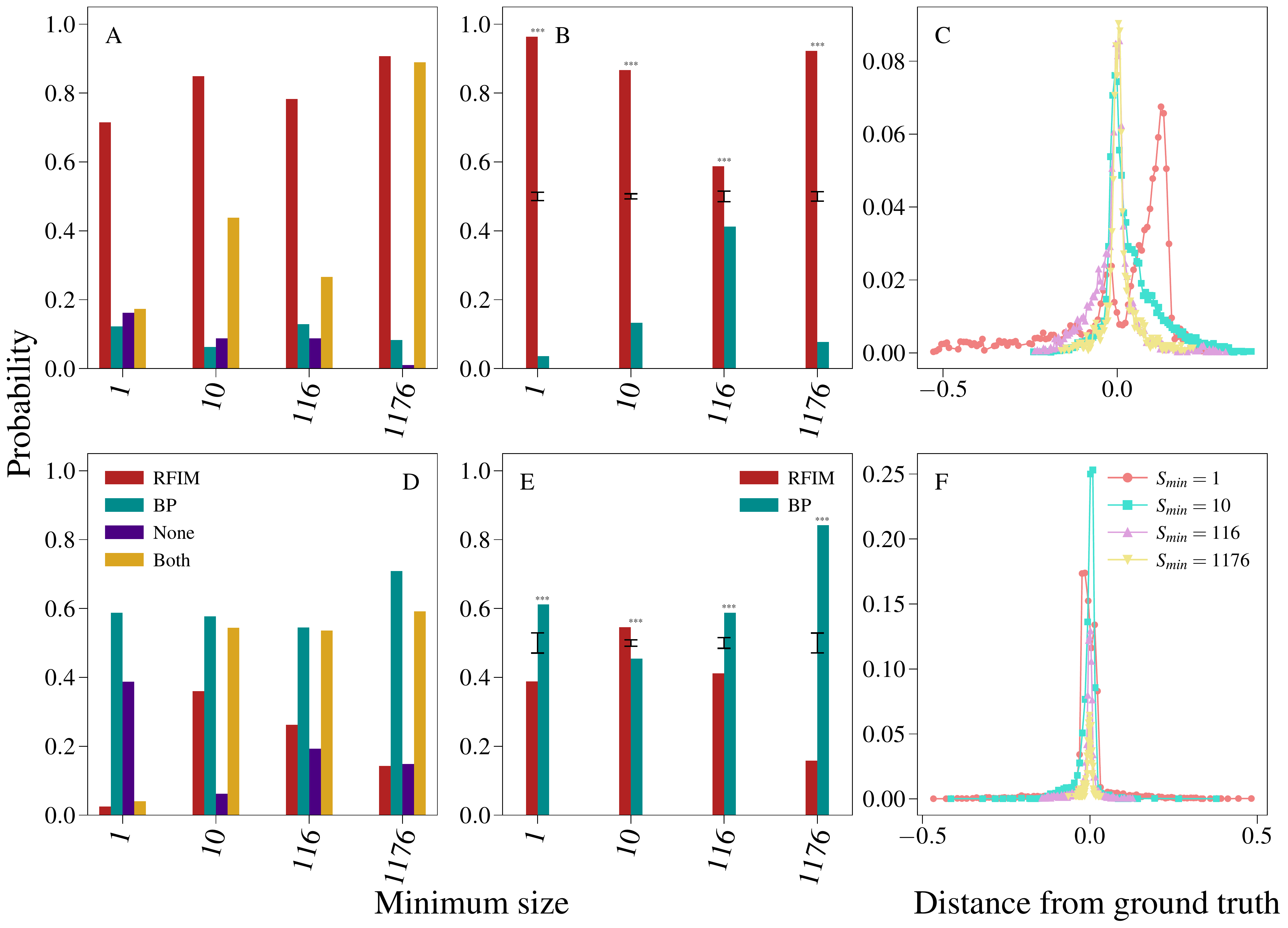}    
\end{center}
\caption{Fitting procedure applied to synthetic data sets. We show the results obtained when the RFIM is the ground truth (upper row) and when the BP is the
ground truth (bottom row), considering four values of $S_{min}$.
Left column: we report the overall probability that the RFIM (red) or the BP 
(blue) is the selected model, the probability that both models are discarded
(purple)  and the probability that both the models are not individually rejected
so that the model selection is performed by means of the log-likelihood ratio test (yellow).
Central column: we report the probability that the RFIM (red) or the BP (blue) is
the model selected by means of the log-likelihood ratio test. Error bars 
represent $\sigma/N$, where $N$ is the sample size and $\sigma =\sqrt{0.25N}$ is
the standard deviation of a binomial distribution with
probability of success equal to $1/2$. Asterisks are used to denote significant 
deviations from the unbiased binomial model, i.e., three asterisks indicate for 
$p < 0.001$.
Right column: we report the probability distribution of the distance between
the true value of the parameter used to generate the distribution $P$ and the 
parameter inferred by fitting against the true model.} 
\label{fig:synt_fits}
\end{figure}

To validate our fitting procedure we apply it on synthetic distributions 
$P$ generated by the RFIM or by the BP. The method must be able to distinguish
effectively between these two models. To this aim, we fix the system size
to be $N=10^6$ and fit $10^4$ realizations of each model. Results are shown in
Fig.~\ref{fig:synt_fits}. 
The
fitting procedure is able to identify the ground truth, either RFIM or BP, regardless of the $S_{min}$ value. In those
cases in which model selection requires the log-likelihood 
ratio test, it still generally holds that the true model is selected
with higher chances. In the case of synthetic data we can also compare
the inferred parameter with the ground truth and Fig.~\ref{fig:synt_fits} C and F
show that the probability that these two quantities differ 
decays quickly as the difference departs from zero. 

\subsection{Robustness of the fits}

\begin{figure}[!htb]
\begin{center}
\includegraphics[width=0.95\textwidth]{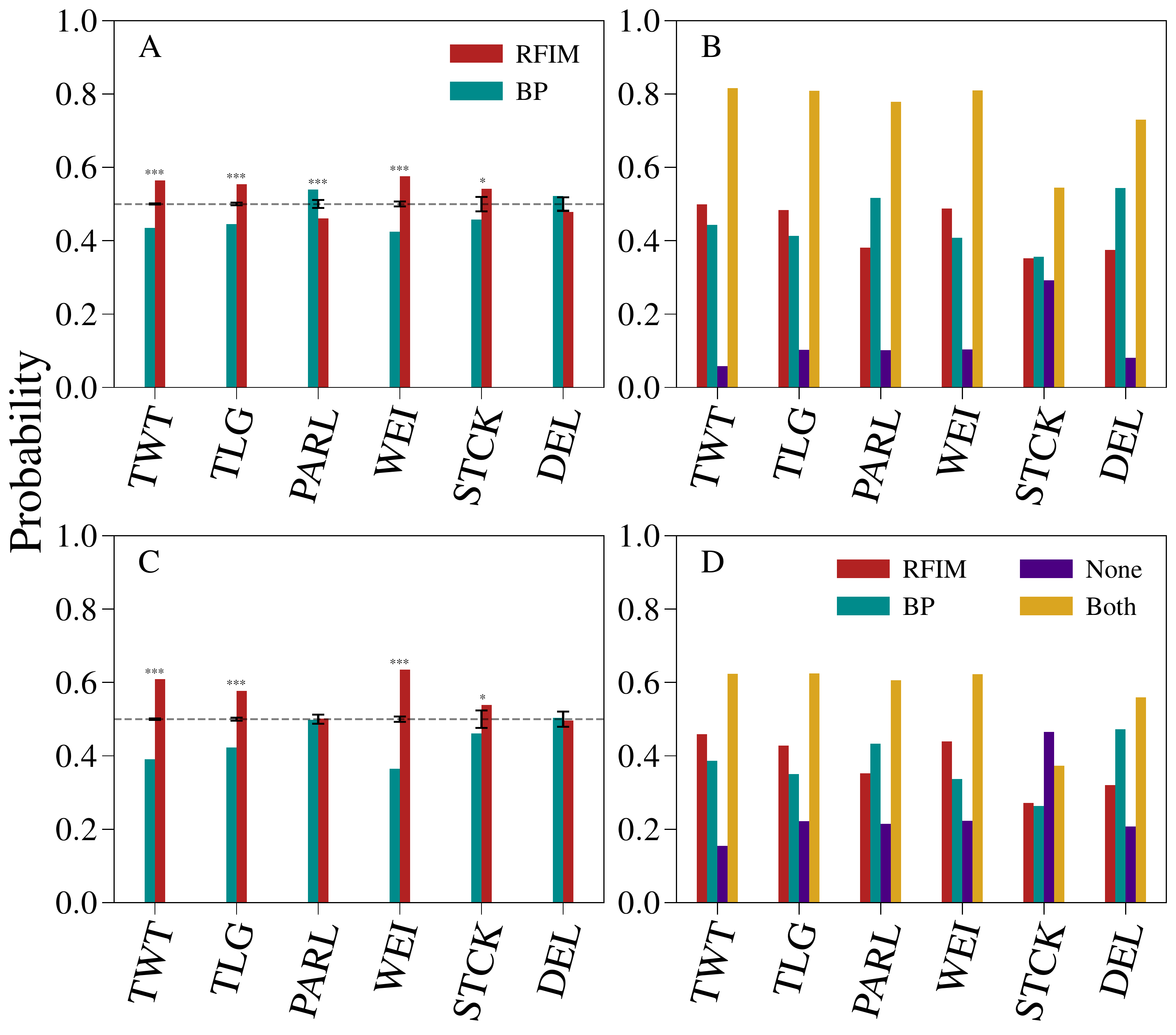}    
\end{center}
\caption{Robustness against the level of statistical significance.
We set 
statistical significance equal to
0.05 (upper row) and 0.2 (lower row).
Acronyms of the data sets are the same used in the main text.
A) Probability that the log-likelihood ratio test favors RFIM over
BP (blue), or vice versa BP over RFIM (red), using a threshold 0.05 over the $p$ values. Only TS that are sufficiently well 
fitted by both models are considered in the analysis, see panel B. 
Error bars represent $\sigma / N$, where $N$
is the sample size and $\sigma = \sqrt{0.25N}$ is the standard deviation of a binomial distribution with
probability of success equal to $1/2$. Asterisks are used to denote significant 
deviations from the unbiased binomial model, i.e., two asterisks indicate for
$p < 0.01$ and one asterisk stands for $p < 0.1$.
B) We report the
fraction of TS that are classified in the RFIM class (red), the fraction of TS 
that are classified as BP (blue), the fraction of TS that is classified as 
neither BP nor RFIM (purple) and the fraction of TS that pass both statistical 
tests (yellow). 
In this case, the log-likelihood ratio test
is required for model selection, see panel A. Here we set to 0.05 the threshold over the $p$ values.
C) Same as in panel A, but the threshold over the $p$ values is set to 0.2.
D) Same as in panel B, but the threshold over the $p$ values is set to 0.2.
} 
\label{fig:fit_varyPval}
\end{figure}

\begin{figure}[!htb]
\begin{center}
\includegraphics[width=0.95\textwidth]{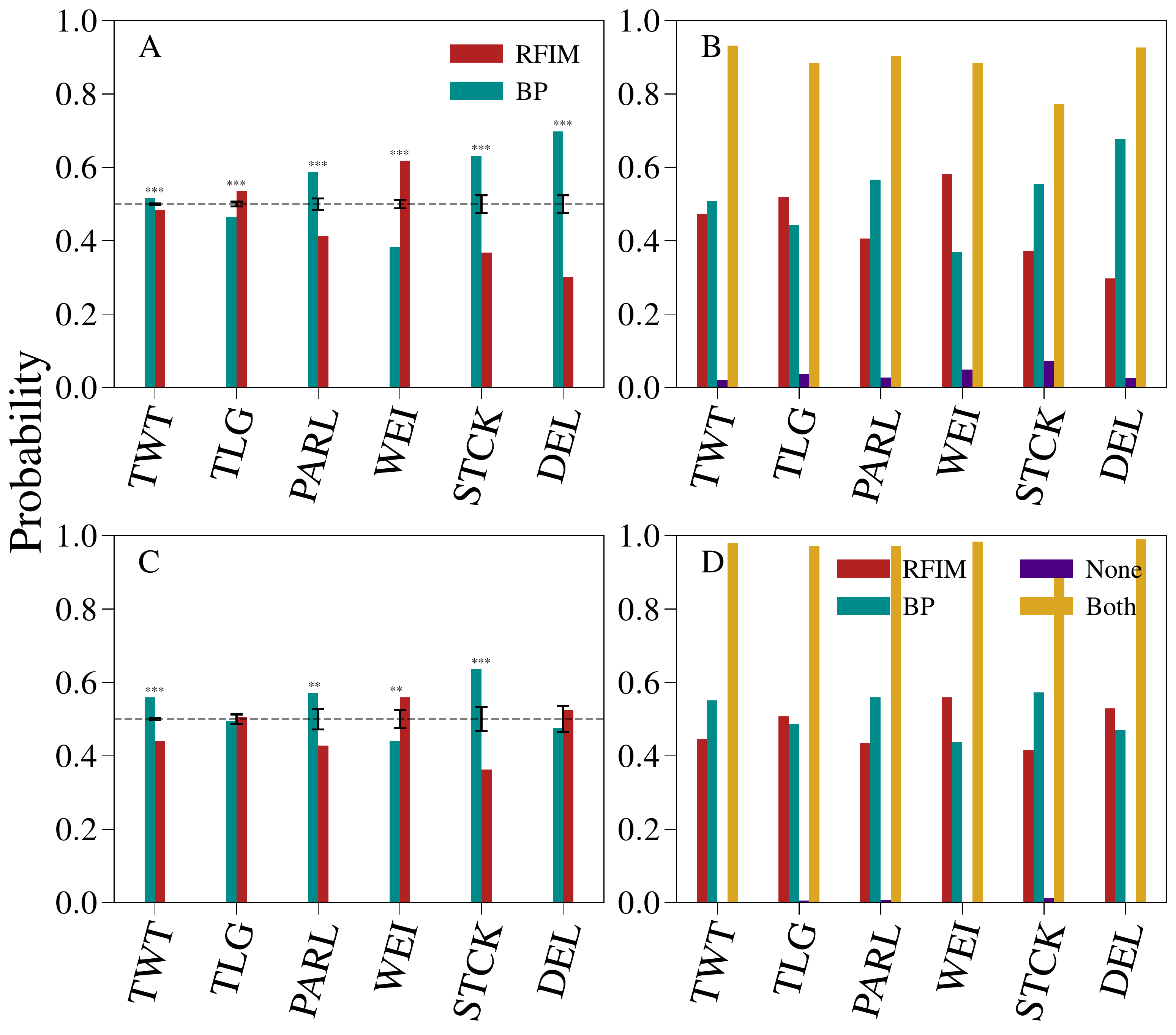}    
\end{center}
\caption{
Robustness against $S_{min}$ values.
Here we use $S_{min}=27$ (upper row) and $S_{min}=95$ (lower row).
Acronyms of the data sets are the same used in the main text.
A) Probability that the log-likelihood ratio test favors RFIM over
BP (blue), or vice versa BP over RFIM (red), using $S_{min}=27$. Only TS that are sufficiently well 
fitted by both models are considered in the analysis, see panel B. 
Error bars represent $\sigma / N$, where $N$
is the sample size and $\sigma = \sqrt{0.25N}$ is the standard deviation of a binomial distribution with
probability of success equal to $1/2$. Asterisks are used to denote significant 
deviations from the unbiased binomial model, i.e., two asterisks indicate for
$p < 0.01$ and one asterisk stands for $p < 0.1$.
B) We report the
fraction of TS that are classified in the RFIM class (red), the fraction of TS 
that are classified as BP (blue), the fraction of TS that is classified as 
neither BP nor RFIM (purple) and the fraction of TS that pass both statistical 
tests (yellow). 
In this case, the log-likelihood ratio test
is required for model selection, see panel A. Here we use $S_{min}=27$.
C) Same as in panel A, but $S_{min}=95$.
D) Same as in panel B, but $S_{min}=95$.
} 
\label{fig:fit_varySmin}
\end{figure}

In the main text we show results of the fitting protocol using 
$S_{min}=10$. Further, we estimated 
statistical significance by setting the threshold value to $0.1$.
Our conclusions, however, are unaffected by different choices of 
these parameters. In Fig.~\ref{fig:fit_varyPval} and 
Fig.~\ref{fig:fit_varySmin},
we vary the threshold over the $p$-value and $S_{min}$,
respectively.

\end{document}